\documentclass[final,onefignum,onetabnum,reqno]{siamart190516}

\usepackage[a4paper,top=2.0cm,bottom=2.0cm,left=2.3cm,right=2.0cm]{geometry}
\usepackage{lipsum}
\usepackage{amsfonts,physics}
\usepackage{graphicx,subfigure}
\usepackage{epstopdf}
\usepackage{algorithmic}  
\usepackage{tikz}
\usepackage{mathtools}
\usepackage{booktabs}
\usepackage{multirow}
\usepackage{siunitx}
\usetikzlibrary{calc}
\usetikzlibrary{backgrounds}
\usepackage{pgfplots,pgfplotstable}
\usetikzlibrary{spy}  
\pgfplotsset{compat=1.5.1}
\pgfplotsset{every axis/.append style={
                    xlabel={$x$},          
                    ylabel={$y$},          
                    label style={font=\small},
                    title style={font=\small},
                    tick label style={font=\tiny}  
                    }}

\usepackage{newfloat}
\ifpdf
  \DeclareGraphicsExtensions{.eps,.pdf,.png,.jpg}
\else
  \DeclareGraphicsExtensions{.eps}
\fi

\newsiamremark{remark}{Remark}
\newsiamremark{hypothesis}{Hypothesis}
\crefname{hypothesis}{Hypothesis}{Hypotheses}
\newsiamthm{claim}{Claim}

\usepackage[framemethod=TikZ]{mdframed}
\newcounter{exam}[section] 
\renewcommand{\theexam}{\arabic{exam}}
\numberwithin{exam}{section}
\crefname{exam}{example}{examples}
\Crefname{exam}{Example}{Examples}

\creflabelformat{exam}{(#2#1#3)}
\newenvironment{exam}[2][]{%
\refstepcounter{exam}%
\ifstrempty{#1}%
{\mdfsetup{%
frametitle={%
\tikz[baseline=(current bounding box.east),outer sep=0pt]
\node[line width=2pt,anchor=east,rectangle,draw=blue!20,fill=blue!20]
{\strut Example~\theexam};}}
}%
{\mdfsetup{%
frametitle={%
\tikz[baseline=(current bounding box.east),outer sep=0pt]
\node[line width=2pt,anchor=east,rectangle,draw=blue!20,fill=blue!20]
{\strut Example~\theexam:~#1};}}%
}%
\mdfsetup{innertopmargin=10pt,linecolor=blue!20,%
linewidth=2pt,topline=true,%
frametitleaboveskip=\dimexpr-\ht\strutbox\relax
}
\begin{mdframed}[]\relax%
\label{#2}}{\end{mdframed}}

\headers{Estimation of spatially varying parameters}{Angwenyi David}

\title{Estimation of spatially varying parameters with application to hyperbolic spdes \thanks{Submitted to the editors on \today.}}

\author{Angwenyi David\thanks{Masinde Muliro University of Science and Technology 
  (\email{dangwenyi@mmust.ac.ke}).}}

\usepackage{amsopn}

\ifpdf
\hypersetup{
  pdftitle={Comparison of the performance of different filters in time-continuous dual estimation},
  pdfauthor={Angwenyi David}
}
\fi

\begin{document}

\maketitle

\begin{abstract}
Parameter estimation is a growing area of interest in statistical signal processing. Some parameters in real-life applications vary in space as opposed to those that are static. Most common methods in estimating parameters involve solving an optimization problem where the cost function is assembled variously; for example, maximum likelihood and maximum a posteriori methods. However, these methods do not have exact solutions to most real-life problems. It is for this reason that Monte Carlo methods are preferred. In this paper, we treat the estimation of parameters which vary with space. We use Metropolis-Hastings algorithm as a selection criteria for the maximum filter likelihood. Comparisons are made with the use of joint estimation of both the spatially varying parameters and the state. We illustrate the procedures employed in this paper by means of two hyperbolic SPDEs: the advection and the wave equation. The Metropolis-Hastings procedure registers better estimates. 
\end{abstract}



\section{Introduction}\label{sec:2.1}
Most state-space models are characterized by, among other things, parameters---which can be constant or varying. By a parameter is comprehended and signified a measurable factor which defines a model and influences its operation. As the parameter changes, so does the model; expressed differently, a parameter is unique to a model it characterises. The choice of a certain model, therefore, is achieved by choosing the right parameters. It occurs more often than not---in hidden Markov models, for instance---that measurements are available but the underlying signal is not readily apparent. This forms an example where parameter estimation is paramount: measurements are used to learn the model parameters, which, in turn, are used to fit the model. 

Consider a signal and measurement equations, with spatially varying parameter signified by a $d-$dimensional vector, $\theta$, 
\begin{subequations}
\begin{align}
\textrm{\textcolor{black}{Signal}:} & \quad dx_{t} = f(x_{t}, \theta)dt  + g(x_{t})d\beta_{t}; \quad t_{0} \leq t, \label{eq:5.1.1a}\\
\textrm{\textcolor{black}{Measurement}:}& \quad dy_{t} = h(x_{t}, \theta)dt + R^{1/2}(t)d\eta_{t};  \quad t_{0} \leq t, \label{eq:5.1.1b}
\end{align}
\end{subequations}
where
\begin{center}
\begin{tabular}{lll}
\textbf{Term} & \textbf{Name} & \textbf{Dimension}\\ 
$x_{t}$ & state vector & $n \times 1$\\ 
$f(x_{t},t)$ & drift function & $n \times 1$ \\ 
$g(x_{t},t)$ & diffusion function & $n \times m$ \\ 
$\{\beta_{t},\, t>t_{0}\}$ & Brownian motion process & $m \times 1$ \\ 
$y_{t}$ & output vector & $r \times 1$ \\ 
$h(x_{t},t)$ & sensor function & $ r \times 1$ \\ 
$R(t)$ & time-function matrix & $r \times r$ \\
$\{\eta_{t},\, t>t_{0}\}$ & standard Brownian motion process & $r \times 1$\\ 
\end{tabular} 
\end{center}

Parameter estimation problem concerns finding the optimal parameter so that the signal best fits the data \cite{Ein,Sar}. This is, classically, achieved by optimization procedure where a cost function is minimized \cite{Lew}. The cost function mostly defines the discrepancy between the state and the measurements. Intuitively, parameter estimation can be seen as a procedure for seeking a parameter value that gives the least discrepancy between the state and the corresponding measurements (also known as the algebraic distance or the residual). The method of least squares has been extensively used to define an objective function. Given the increment in measurement, $dy_{t}$, of the state, $x_{t}$, at time $t$, the objective function, $\mathcal{J}(\theta)$, in the least squares sense, is given by
\begin{equation}
 \mathcal{J}(\theta) = \int_{t_{0}}^{t}w_{t}\Vert dy_{t} - h(x_{t}, \theta ) dt\Vert^{2},
 \end{equation} 
where $w_{t}$ is the weighting function.

Most commonly used procedures in the framework of least-squares include: linear least-squares, orthogonal least-squares, gradient weighted least-squares, bias corrected renormalization. This paper, however, attends not to the study of least-squares approaches, they being outside the scope of its design. Suffice it to only direct the interested reader to the article \cite{Zha} for an elaborate explanation and application of least-squares methods in computer vision. Instead, we study parameter estimation by means of filtering. But before that, we mention a few merits and demerits of least-squares and other methods defined by algebraic distances. 

The use of algebraic distances in defining a cost function is computationally efficient and closed-form solutions are possible. The end result, however, is not satisfactory. This is due, in one part, to the fact that the objective function is mostly not invariant with respect to Euclidean transformations, for example, translation. This limits the coordinates systems to be used. In the other part, outliers may not contribute to the parameters the same way as inliers \cite{Zha}. Other more satisfactory parameter estimation methodologies are highly desired. We consider, in this paper, the use of Bayesian inference techniques.

Estimation of parameters by means of a filter can be achieved in a number of ways; one of them being the use of the filter evidence, or its near approximation, and selection criteria for parameters which give a reasonable estimate of the evidence. The second method involves updating the parameters and the state at the same time. This is known as dual estimation, which further subdivides into joint estimation and a dual filter. Joint estimation entails subjoining the vector of parameters to the state vector to form an extended state-space. The filter is then implemented and run forward in time with the hope of filter convergence to the optimal state and parameter values. A dual filter, on the other hand, involves implementing a filter for the state and that of parameters simultaneously. The filter provides a self-correcting mechanism which may lead to convergence of state and parameter estimates. 

The rest of this paper is arranged as follows. In the first place, we introduce some notions on Bayesian inference of parameters where we pass the limit in the mean in order to move from a discrete setting to a continuum in time. We introduce the different ways in which a filter can be used for parameter estimation. In anticipation of application in the later part of the paper, we introduce the stochastic advection and wave equation together with their spatial discretization. We then illustrate how parameters are to be estimated by means of a filter likelihood and the dual filters. Results and discussions follow and the conclusion forms the closing part of this paper. 

\section{Bayesian parameter inference}
In Bayesian inference of parameters, the parameters are treated as a random variable. The parameter is assigned a prior, $\pi_{t_{0}}(\theta)$, based on some initial belief. Let $t_{n}$ such that $t_{n+1}>t_{n} \; \forall n = 0,\, 1, \, 2, \, ...\, N $ be a partition of the interval $[t_{0}, T]$ and let $\delta t= t_{n+1}-t_{n}$. Bayes' rule gives the joint posterior of parameters and the states; 
\begin{subequations}
\begin{align}
\begin{split}
\pi_{[t_{0},T]}(x,\, \theta \mid Y_{T}) & \approx \underset{N \to \infty}{\underset{\delta t\to 0}{\textrm{l.i.m. }}} \pi_{t_{0}:t_{N}}(x_{t_{0}:t_{N}},\, \theta \mid y_{t_{0}:t_{N}}) \\
& = \underset{N \to \infty}{\underset{\delta t\to 0}{\textrm{l.i.m. }}} \dfrac{\pi_{t_{0}:t_{N}}(y_{t_{0}:t_{N}} \mid x_{t_{0}:t_{N}},\, \theta)\pi_{t_{0}:t_{N}}(x_{t_{0}:t_{N}} \mid \theta) \pi_{t_{0}}(\theta)}{\pi_{t_{0}:t_{N}}(y_{t_{0}:t_{N}})}, \label{5.2.1a}
\end{split}
\intertext{where $Y_{T}=y_{[t_{0},T]}$, }
\pi_{t_{0}:t_{N}}(y_{t_{0}:t_{N}} \mid x_{t_{0}:t_{N}},\, \theta) &= \prod_{n=1}^{N}\pi_{t_{n}}(y_{t_{n}} \mid x_{t_{n}},\, \theta), 
\intertext{and}
\pi_{t_{0}:t_{N}}(x_{t_{0}:t_{N}} \mid \theta) & = \pi_{t_{0}}(x_{t_{0}} \mid \theta)\prod_{n=1}^{N}\pi_{t_{n}}(x_{t_{n}} \mid x_{t_{n-1}}, \, \theta).
\end{align}
\end{subequations}
Now to arrive at the marginal posterior of parameters, we integrate out the states from the joint posterior of states and parameters, \cref{5.2.1a}:
\begin{equation}
\pi_{t_{0}:t_{N}}(\theta \mid y_{t_{0}:t_{N}})  = \int \dfrac{\pi_{t_{0}:t_{N}}(y_{t_{0}:t_{N}} \mid x_{t_{0}:t_{N}},\, \theta)\pi_{t_{0}:t_{N}}(x_{t_{0}:t_{N}} \mid \theta) \pi_{t_{0}}(\theta)}{\pi_{t_{0}:t_{N}}(y_{t_{0}:t_{N}})}dx_{t_{0}:t_{N}}. \label{5.2.2}
\end{equation}
It turns out that direct computation of the integral in \cref{5.2.2} is difficult, especially with the increase in measurements \cite{Sar}. This challenge is circumvented through the use of recursive techniques which include the use of filters and smoothers, maximum a posteriori (MAP) estimates, and drawing samples from the posterior using Markov Chain Monte Carlo (MCMC) methods. 

To use recursive methods, we begin with the following parameter posterior
\begin{subequations}
\begin{align}
\pi (\theta \mid Y_{T} ) & \approx \underset{N \to \infty}{\underset{\delta t\to 0}{\textrm{l.i.m. }}} \pi (\theta \mid y_{t_{0}:t_{N}} ) \propto \underset{N \to \infty}{\underset{\delta t\to 0}{\textrm{l.i.m. }}} \pi_{t_{0}:t_{N}} (y_{t_{0}:t_{N}} \mid \theta)\pi_{t_{0}} (\theta).
\intertext{where}
\begin{split}
\pi_{t_{0}:t_{N}} (y_{t_{0}:t_{N}} \mid \theta) & = \prod_{n=1}^{N} \pi_{t_{n}}(y_{t_{n}} \mid y_{t_{1}:t_{n-1}}, \, \theta) \\
												& = \prod_{n=1}^{N} \pi_{t_{n}}(y_{t_{n}} \mid x_{t_{n}}, \, \theta) \pi_{t_{n}}(x_{t_{n}} \mid y_{t_{1}:t_{n-1}}, \, \theta)dx_{t_{n}}
\end{split}
\end{align}
\end{subequations}

The state's predictive distribution, $\pi_{t_{n}}(x_{t_{n}} \mid y_{t_{1}:t_{n-1}}, \, \theta)$, can be computed recursively as follows \cite{Sar}:
\begin{subequations}
\begin{align}
\pi_{t_{n}}(x_{t_{n}} \mid y_{t_{1}:t_{n-1}}, \, \theta) = \int \pi_{t_{n}}(x_{t_{n}} \mid x_{t_{n-1}}, \, \theta) \pi_{t_{n-1}}(x_{t_{n-1}} \mid y_{t_{1}:t_{n-1}}, \, \theta)dx_{t_{n-1}}.
\end{align}
\end{subequations}

Instead of working with the posterior, $\pi (\theta \mid Y_{T} )$, it is quite convenient to use the negative log-posterior obtained by expressing the posterior as follows
\begin{equation}
\pi (\theta \mid Y_{T} )  \approx \underset{N \to \infty}{\underset{\delta t\to 0}{\textrm{l.i.m. }}} \pi (\theta \mid y_{t_{0}:t_{N}} ) \propto \underset{N \to \infty}{\underset{\delta t\to 0}{\textrm{l.i.m. }}}\exp (-\psi_{T}(\theta)),
\end{equation}
where the \textit{energy function}, $\psi_{T}(\theta)$, is given by
\begin{equation}
\psi_{T}(\theta) = - \log (\pi_{t_{0}:t_{N}} (y_{t_{0}:t_{N}} \mid \theta)) - \log (\pi_{t_{0}}(\theta)).
\end{equation}

The maximum a posteriori estimate (MAP) can then be obtained by the mode of the posterior distribution, or, equivalently, the minimum of the energy function, the latter of which is easier to compute; that is,
\begin{equation}
\begin{split}
\hat{\theta}_{\textrm{MAP}} &= \underset{\theta}{\textrm{arg max }} \pi (\theta \mid Y_{T} ) \\
											& = \underset{\theta}{\textrm{arg min }} \psi_{T}(\theta). \label{eq:6.2.7}
\end{split}
\end{equation}
One demerit of MAP estimate is that it yields a point estimate of the parameter posterior, and therefore ignores the dispersion of the estimate. Setting the prior, $\pi_{t_{0}}(\theta)$, to be a uniform density then \cref{eq:6.2.7} yields a Maximum Likelihood estimate.

\section{Metropolis-Hastings method} 
Metropolis-Hastings\footnote{Named after Nicholas Constantine Metropolis (1915-1999) and Wilfred Keith Hastings (1930-2016)} \cite{Rob} is a Markov-Chain Monte Carlo sampling algorithm with optimal convergence. It is premised on detailed balance and ergodicity. Given a probability density, say, $\pi(\theta)$, from which it is difficult to sample---for instance if the said distribution is known to a normalisation constant---, and given another distribution $\rho(\theta)$, say, from which it is easy to sample, then detailed balance is the condition
\begin{equation}
\pi(\theta_{k}) \rho(\theta_{k} \mid \theta_{k+1}) = \pi(\theta_{k+1}) \rho(\theta_{k+1}\mid \theta_{k}),
\end{equation}
where $\rho(\theta_{k}\mid \theta_{k+1})$ is a transition distribution. The detailed balance condition is necessary for any random walk to asymptotically converge to a stationary distribution. By ergodicity is meant that there is a non-zero probability in moving from a state to any other state in a Markov-Chain.

The following algorithm summarises the Metropolis-Hastings procedure

\begin{algorithm}[htp]
\caption{Metropolis-Hastings}
\label{alg:05.1}
\begin{algorithmic}[1]
\STATE{Draw $\tilde{\theta}$ from  $\rho(\tilde{\theta}\mid \theta_{k})$}
\STATE{Set $\theta_{k+1} \leftarrow \tilde{\theta}$ with probability $\alpha = \min \left( 1, \, \dfrac{\pi(\tilde{\theta}) \rho(\theta_{k}\mid \tilde{\theta}) }{\pi(\theta_{k}) \rho(\tilde{\theta}\mid \theta_{k}) }\right) $}
\STATE{Otherwise set $\theta_{k+1} \leftarrow \theta_{k}$ }
\end{algorithmic} 
\end{algorithm}

\section{Dual estimation}
Dual estimation comprehends simultaneous estimation of state and parameters by means of an appropriate filter. The self-correcting mechanism of the filter is taken advantage of to converge to both the true state and the true parameters. Depending on the initial parameter, the filter sooner or later converges to the true parameter value. Dual estimation can be achieved in two ways: joint estimation and by a dual filter \cite{Lu,Lint,Mora}. 
\subsection{Joint estimation (augmented state-space)}
In joint estimation, the state vector is augmented with the vector of parameters to form an extended state-space and then the filter is run forward in time for an update of both the state and the parameters. The parameters are induced with artificial dynamics, or are made to assume a random walk; that is, respectively,
\begin{equation}
 dz_{t} = \zeta_{t} ; \quad t_{0} \leq t, \label{eq:6.4.1}
\end{equation}
where 
\begin{subequations}
\begin{align}
dz_{t} & =\begin{pmatrix}
dx_{t} \\
d\theta_{t}
\end{pmatrix} \textrm{  and  } \zeta_{t}=\begin{pmatrix}
f(x_{t},\theta)dt  + g(x_{t})d\beta_{t} \\
0
\end{pmatrix},
\intertext{or where}
\zeta_{t} & =\begin{pmatrix}
f(x_{t},\theta)dt  + g(x_{t})d\beta_{t} \\
\sigma d\chi_{t}
\end{pmatrix},
\end{align}
\end{subequations}
in which $\{\chi_{t},\, t>t_{0}\}$ is a $d-$dimensional standard Brownian motion vector process and $\sigma$ is a small constant. A filter is then implemented with the augmented state $z_{t}$ in the place of $x_{t}$. The demerit of this method is that the extended state-space has an increased degree of freedom owing to many unknowns, of both the state and the parameters, which renders the filter unstable and intractable, especially in nonlinear models \cite{Mor}.

\subsection{Dual filter} \label{subsec:5.3.2}
Dual filtering of the state and parameters is attained by use of two filters, one for state update and another for updating parameters, both run simultaneously. The two filters interact symbiotically in that one provides the update of the state to be used by the other, while the other provides an update of the parameters to be used by the former. A very good example in literature is the dual extended Kalman filter \cite{Wan} used for estimating neural network models and the weights. In this case, the state is the model signal and the weights are parameters. Another example appears in \cite{Ang} where a dual filter comprising of the ensemble transform particle filter (ETPF) and the feedback particle filter (FPF) is used for simultaneous estimation of the state of a wave equation and its speed. 

The model for the dual filter of our consideration comprises of a $d-$dimensional vector equation of artificial dynamics of parameters together with the state space model, \Cref{eq:5.1.1a,eq:5.1.1b}: 
 \begin{subequations}
\begin{align}
\textrm{\textcolor{red}{Parameter}:} & \quad d\theta_{t} = 0, \qquad \qquad t_{0} \leq t, \label{eq:5.3.2a}\\
\textrm{\textcolor{red}{Signal}:} & \quad dx_{t} = f(x_{t}, \theta_{t})dt  + g(x_{t})d\beta_{t}, \quad t_{0} \leq t, \label{eq:5.3.2b}\\
\textrm{\textcolor{red}{Measurement}:}& \quad dy_{t} = h(x_{t})dt + R^{1/2}(t)d\eta_{t},  \quad t_{0} \leq t, \label{eq:5.3.2c}
\end{align}
\end{subequations}
where the nomenclature and dimensions remain as stipulated for \Cref{eq:5.1.1a,eq:5.1.1b}. 

In the following, we introduce the equations to which we shall apply dual filters in estimating spatially varying parameters.

\section{Application equations}
\subsection{Advection Equation} \label{subsec:7.1.1}
We take up an advection equation, excited with space-time white noise process, with some diffusion term added to it, and on a periodic spatial domain of length $L$, which equation we write as follows: 
\begin{equation}
\dfrac{\partial u}{\partial t} = \dfrac{\partial (C(x) u)}{\partial x}   + \mu \dfrac{\partial^{2} u}{\partial x^{2}} + \sigma \dot{\beta}_{x,t}, \; \; \; 0\leq t \leq T_{t} \times 0\leq x \leq L, \label{eq:12a}
\end{equation}
where $C(x)$ is a spatially varying velocity (of which constant velocity is a special case), $\sigma$ is a constant, and $u(x,t)$ is the function to be obtained, which function describes the state of the signal. $\mu$ is a constant whilst $\dot{\beta}_{x,t}$ is space-time white noise process where the dot denotes the singularity of the noise process. 

\Cref{eq:12a} needs a remark owing to the roughness of the stochastic forcing term $\dot{\beta}_{x,t}$, which is a mixed distributional derivative of Brownian sheet. As is well known (see \cite{Stu} for details), the Brownian sheet is nowhere differentiable. We, however, use the method introduced in \cite{All}; that is, we approximate the noise term as follows. Let the domain $0\leq t \leq T_{t} \times 0\leq x \leq L$ be tessellated into rectangles $[t_{n},t_{n+1}] \times [x_{i}, x_{i+1}]$ of dimensions $\delta t \times \delta x$ each, for $n=1,\; 2,\;3,\;...,\; T$ and $i=1,\; 2,\;3,\;...,\; N$ so that $\delta t = T_{t}/T$ and $\delta x = L/N$. Then,
\begin{equation}
 \dot{\beta}_{x,t} \approx \dfrac{1}{\delta x \delta t}\sum_{i=1}^{N}\sum_{n=1}^{T}\omega_{i,n}(\delta x \delta t)^{1/2}\chi_{i}(x)\chi_{n}(t),
 \end{equation} 
 where $\{\omega_{i,n}\}_{i=1}^{N}$ are independent and identically distributed random variables of mean $0$ and unit variance. $\chi_{i}(x)$ and $\chi_{n}(t)$ are characteristic functions, and are given by
 \[\chi_{n}(t)=\begin{cases}
    1,& \text{if } t_{n}\leq t \leq t_{n+1},\\
    0,              & \text{otherwise},
\end{cases} \]
  \[\chi_{i}(x)= \begin{cases}
    1,& \text{if } x_{i}\leq x \leq x_{i+1},\\
    0,              & \text{otherwise}.
\end{cases} \]

By three-point upwind scheme in space \cite{Cou}, we  discretise \cref{eq:12a} and arrive at the following
\begin{equation}
\begin{split}
\dfrac{du_{i}}{dt} & \approx \dfrac{3C_{i} u_{i}- 4C_{i-1}u_{i-1}+C_{i-2}u_{i-2}}{2\delta x} \\
& \quad{} + \mu \dfrac{3u_{i+2}-16u_{i+1} +26u_{i}-16u_{i-1}+3u_{i-2} }{4\delta x^{2}} + \sigma \dfrac{1}{\sqrt{\delta x}} \dot{\omega}_{i}, \label{eq:12b}
\end{split}
\end{equation}
where $\delta x$ is the spatial step size and $\{\omega_{i,t}, t>t_{0}\}$ is standard Brownian motion process. The $i$th grid point is represented by $ x_{i}=i \delta x$. With this notation, $u_{i,n}$ is understood to mean the value of $u$ at the $i$th grid point at time $t_{n}$. 

Time discretisation, by means of Euler-Maruyama scheme, leads to 
\begin{equation}
\begin{split}
u_{i,t_{n+1}} & = u_{i,t_{n}}+  \dfrac{3C_{i} u_{i,t_{n}}- 4C_{i-1}u_{i-1,t_{n}}+C_{i-2}u_{i-2,t_{n}}}{2\delta x}\delta t \\
 & \quad {} + \mu \dfrac{3u_{i+2,t_{n}}-16u_{i+1,t_{n}} +26u_{i,t_{n}}-16u_{i-1,t_{n}}+3u_{i-2,t_{n}} }{4\delta x^{2}}\delta t + \sigma \dfrac{\delta t^{1/2}}{\sqrt{\delta x}} \omega_{i,t_{n}}, \label{eq:12bc}
\end{split}
\end{equation}
where $\omega_{i,t_{n}}$ is a random variable of mean $0$ and variance $1$. The time increment, $\delta t>0$, is such that the limit of $\delta u_{i}$ as $\delta t\rightarrow 0$ is $du_{i}$. Furthermore, $n=1,\; 2,\;3,\;...,\; T$.
We use the following initial value.
\begin{equation}
u(x,t_{0}) = \sin(x). \label{eq:2.4}
\end{equation}
Moreover, $C(x)=e^{\lambda(x)}$ where $$\lambda(x)=\sin(2\pi x).$$

Considering every grid point in \cref{eq:12bc} leads to a vector representation of the signal $u$. To do so requires the following shorthand for operations,   
\[(D_{1} u)_{i}:=\dfrac{3u_{i}-4u_{i-1}+u_{i-2}}{2\delta x},\;\;\; \qquad \forall i=1,\;2,\;3,\;...,\; N,\]
and
\[(D^{\textrm{T}}_{1} u)_{i}:=\dfrac{u_{i+2}-4u_{i+1}+3u_{i}}{2\delta x},\;\;\; \qquad \forall i=1,\;2,\;3,\;...,\; N,\]
so that 
\begin{equation}
\begin{split}
(D_{1}D^{\textrm{T}}_{1} u)_{i}& :=\dfrac{3u_{i+2}-12u_{i+1}+9u_{i}}{4\delta x^{2}}    + \dfrac{-4u_{i+1}+16u_{i}-12u_{i-1}}{4\delta x^{2}} \\
 & \quad{} + \dfrac{u_{i}-4u_{i-1}+3u_{i-2}}{4\delta x^{2}} \\
 &  = \dfrac{3u_{i+2}-16u_{i+1} +26u_{i}-16u_{i-1}+3u_{i-2}}{4\delta x^{2}}  ,\;\;\; \qquad \forall i=1,\;2,\;3,\;...,\; N.
 \end{split}
\end{equation}
We finally have
\begin{equation}
u_{t_{n+1}}=u_{t_{n}}+F(t_{n}) u_{t_{n}}\delta t + G(t_{n})\omega_{t_{n}},  \label{eq:12bd}
\end{equation}
where $u_{t_{n}}$ is an $N-$dimensional column vector at time $t_{n}$ comprising of elements $u_{i,t_{n}}, \; i=1,\;2,\;3,\;...,\; N$ and
\[
F(t_{n})
=
\begin{bmatrix}
  D_{1}C_{\textrm{diag}} - \mu D_{1} D^{\textrm{T}}_{1}
\end{bmatrix}, \;
G(t_{n})
=
\begin{bmatrix}
\sigma \dfrac{\delta t^{1/2}}{\sqrt{\delta x}} I_{N\times N} 
\end{bmatrix},
\]
in which $C_{\textrm{diag}}$ is a diagonal matrix made of the elements of $C$, and $ I_{N\times N} $ is the $N$th order identity matrix.

The surface and contour plots for the stochastic advection equation are shown below; that is, when $\sigma = 0.1$. 
\begin{figure}[htp]
\centering
\subfigure[Contour plot for the solution to the deterministic advection equation][]{
\includegraphics[width=0.4\textwidth]{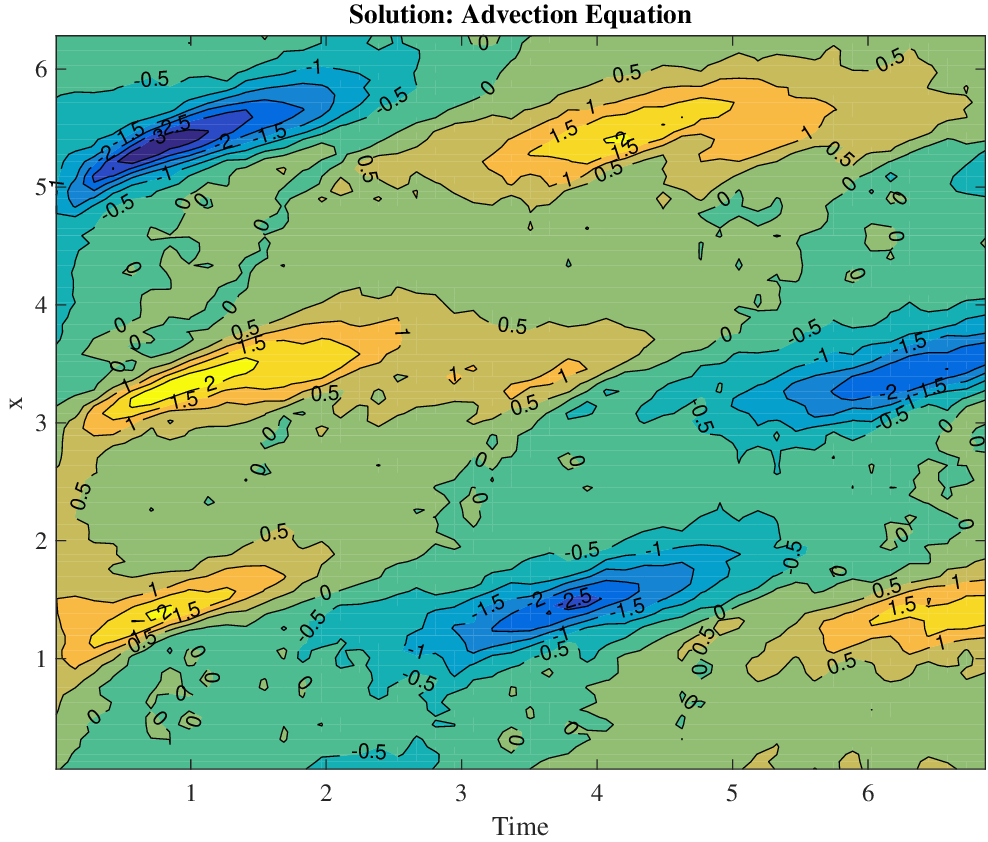}
\label{fig:subfig advr1}}
\subfigure[Surface plot for the solution to the deterministic advection equation][ ]{
\includegraphics[width=0.45\textwidth]{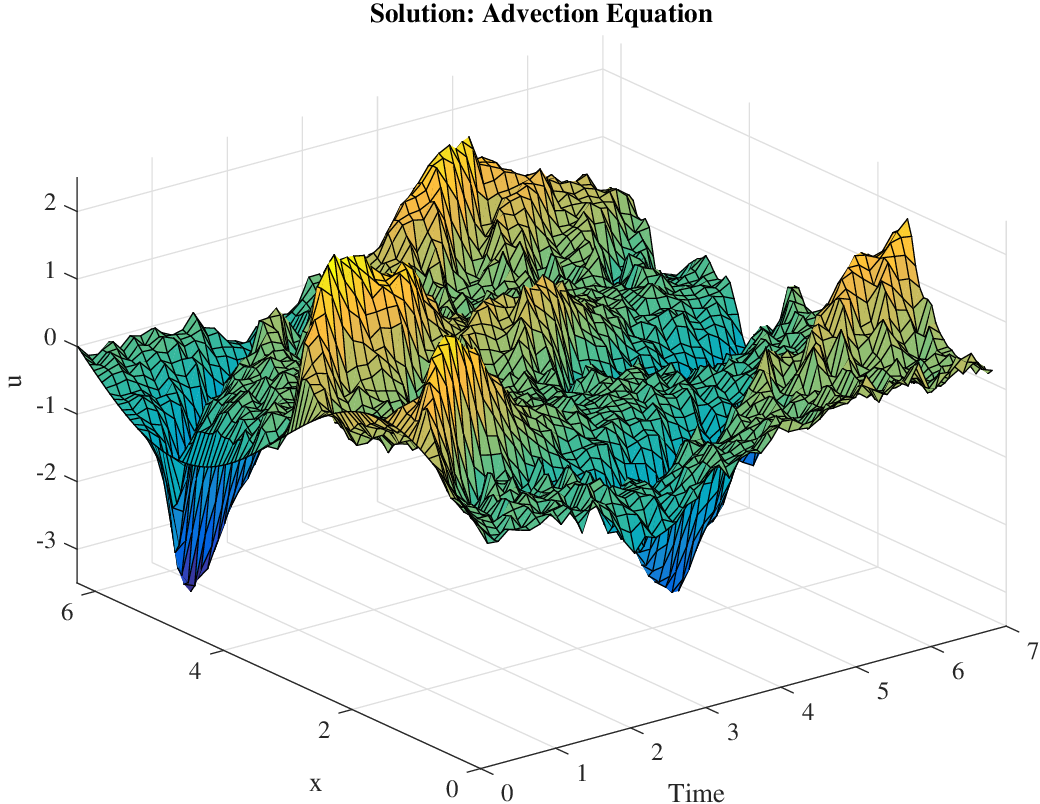}
\label{fig:subfig advr2}}
\caption[Contour and surface plots for one realisation of stochastic advection equation]{Contour and surface plots for a single realisation of the stochastic advection equation under the following setting: $L=2\pi$, $N=100$, $\delta x=L/N$, $\delta t=0.007$, $T=1000$, $\sigma = 0.1$, $\mu=0.01$, $C(x)=e^{\lambda(x)}$ where $\lambda(x)=\sin(2\pi x)$, and $ u_{t_{0}}=\sin(x)$.}
\label{fig:globfig advr7}
\end{figure}
The ruggedness evident in \Cref{fig:globfig advr7} is consequent upon the addition of noise to the underlying dynamics. 

In the next subsection, we introduce the wave equation. 

\subsection{Wave Equation} \label{subsec:7.3.2}

The wave equation---for our consideration---is given by:
\begin{equation}
\dfrac{\partial^{2} u}{\partial t^{2}} = \dfrac{\partial(C(x)\partial u/\partial x)}{\partial x} + \mu \dfrac{\partial^{3} u}{\partial x^{2}\partial t} + \sigma \dot{\beta}_{x,t},  \; \; \; 0\leq t \leq T_{t} \times 0\leq x \leq L, \label{eq:7.1.7}
\end{equation}
 where  $C(x)=e^{\lambda(x)}$ is the wave velocity---and is for a wave travelling in a heterogeneous medium---, and $u(x,t)$ is the function to be obtained, which function, as in the previous subsection, describes the state of the signal. Moreover, $\sigma$ is a constant and $\dot{\beta}_{x,t}$, as before, is the space-time white noise.

We employ mixed difference schemes to discretise \cref{eq:7.1.7} in space, so that we have:
\begin{subequations}
 \begin{align}
 \dfrac{du_{i}}{dt}& \approx  p_{i}, \label{eq:6.2.5a}\\
\dfrac{dp_{i}}{dt}  & \approx \dfrac{C_{i+1} w_{i+1}- C_{i}w_{i}}{\delta x} + \mu \dfrac{p_{i+1} - 2p_{i}+p_{i-1} }{\delta x^{2}} + \dfrac{1}{\sqrt{\delta x}} \dot{\omega}_{i},\label{eq:6.2.5b}
 \end{align}
\end{subequations}
where $w_{i}:=\dfrac{ u_{i} - u_{i-1}}{\delta x}$, and $\delta x$ is the spatial step size. For time integration we use Verlet's method, because of its geometric properties; namely, volume preservation, symplecticity, conservation of first integrals and reversibility \cite{Hai,Rei2}---which method, applied to the deterministic part of \cref{eq:6.2.5a,eq:6.2.5b}, yields
\begin{subequations}
 \begin{align}
 u_{i,t_{n+1}}& = u_{i,t_{n}}+  p_{i,t_{n+1/2}}\delta t, \label{eq:6.2.6a}\\
 p_{i,t_{n+1/2}}  &=p_{i,t_{n}} + \dfrac{C_{i+1} w_{i+1,t_{n}}- C_{i}w_{i,t_{n}}}{\delta x}\dfrac{\delta t}{2} + \mu \dfrac{p_{i+1,t_{n}} - 2p_{i,t_{n}}+p_{i-1,t_{n}} }{\delta x^{2}}\dfrac{\delta t}{2}, \label{eq:6.2.6b}\\ 
 \begin{split}
 p_{i,t_{n+1}}  &=p_{i,t_{n+1/2}} + \dfrac{C_{i+1} w_{i+1,t_{n+1}}- C_{i}w_{i,t_{n+1}}}{\delta x}\dfrac{\delta t}{2} \\
 & \quad {} + \mu \dfrac{p_{i+1,t_{n+1}} - 2p_{i,t_{n+1}}+p_{i-1,t_{n+1}} }{\delta x^{2}}\dfrac{\delta t}{2}+ \sigma \dfrac{\delta t^{1/2}}{\sqrt{\delta x}} \omega_{i,t_{n}}, \label{eq:6.2.6c}
 \end{split}
 \end{align}
 \end{subequations}
 where $\delta t $ is the time step and $w_{i,t_{n}}:=\dfrac{ u_{i,t_{n}} - u_{i-1,t_{n}}}{\delta x}$. Substituting \cref{eq:6.2.6b} into \cref{eq:6.2.6a,eq:6.2.6c}, we get
 \begin{subequations}
 \begin{align}
 \begin{split}
 u_{i,t_{n+1}}& = u_{i,t_{n}}+  p_{i,t_{n}} + \dfrac{C_{i+1} w_{i+1,t_{n}}- C_{i}w_{i,t_{n}}}{\delta x}\dfrac{\delta t}{2} \\
 &\quad{} + \mu \dfrac{p_{i+1,t_{n}} - 2p_{i,t_{n}}+p_{i-1,t_{n}} }{\delta x^{2}}\dfrac{\delta t}{2}, 
 \end{split}\label{eq:6.2.7a}\\
 \begin{split}
 p_{i,t_{n+1}}  &=p_{i,t_{n}} + \dfrac{C_{i+1} w_{i+1,t_{n}}- C_{i}w_{i,t_{n}}}{\delta x}\dfrac{\delta t}{2} \\
 & \quad {} + \mu \dfrac{p_{i+1,t_{n}} - 2p_{i,t_{n}}+p_{i-1,t_{n}} }{\delta x^{2}}\dfrac{\delta t}{2} + \dfrac{C_{i+1} w_{i+1,t_{n+1}}- C_{i}w_{i,t_{n+1}}}{\delta x}\dfrac{\delta t}{2} \\
 & \quad{}  + \mu \dfrac{p_{i+1,t_{n+1}} - 2p_{i,t_{n+1}}+p_{i-1,t_{n+1}} }{\delta x^{2}}\dfrac{\delta t}{2}+ \sigma \dfrac{\delta t^{1/2}}{\sqrt{\delta x}} \omega_{i,t_{n}}. \label{eq:6.2.7b}
 \end{split}
 \end{align}
 \end{subequations}
 We use the following initial values:
\begin{subequations}
 \begin{align}
u(x,0) &= \exp(-4(x-0.5L).^{2}) ,  \label{eq:7.1.11a}\\
p(x,0)  &=0, \label{eq:7.1.11b}
 \end{align}
\end{subequations}
 where $L$ is the length of the domain. Now $C(x)=e^{\lambda(x)}$ where $$\lambda(x)=\sin(x).$$
 Considering every grid point leads to a vector representation of the signal $u$. We use the following shorthand,   
\[(D_{2} u)_{i}:=\dfrac{u_{i}-u_{i-1}}{\delta x},\;\qquad \;\; \forall i=1,\;2,\;3,\;...,\; N.\]
\Cref{eq:6.2.7a,eq:6.2.7b} then become
\begin{equation}
\underline{u}_{t_{n+1}}
= \underline{u}_{t_{n}} +
F{t_{n}}\underline{u}_{t_{n}}
+
G(t_{n})\underline{\omega}_{t_{n}}, \label{eq:7.1.13}
\end{equation}
where
\begin{equation*}
\begin{split}
F(t_{n}) &= - I_{2N\times 2N} +
\begin{bmatrix}
  I_{N\times N}-\dfrac{\delta t^{2}}{2} D_{2}(C_{\textrm{diag}}(x) D^{\textrm{T}}_{2} &  \delta t I_{N\times N} - \dfrac{\delta t^{2}}{2}\mu D_{2} D^{\textrm{T}}_{2} \\
   -\dfrac{\delta t}{2} D_{2}(C_{\textrm{diag}}(x) D^{\textrm{T}}_{2}  &  I_{N\times N} - \dfrac{\delta t}{2}\mu D_{2} D^{\textrm{T}}_{2}
\end{bmatrix} \\
& \quad {} \times
\begin{bmatrix}
  I_{N\times N} & 0_{N\times N}\\
   \dfrac{\delta t }{2}D_{2}(C_{\textrm{diag}}(x) D^{\textrm{T}}_{2})  &   I_{N\times N} - \dfrac{\delta t}{2}\mu D_{2} D^{\textrm{T}}_{2}
\end{bmatrix}^{-1}, 
\end{split}
\end{equation*}
\[
\underline{u}=
\begin{pmatrix}
p\\
u
\end{pmatrix}, \;
G(t_{n})
=
\begin{bmatrix}
0_{N\times N} & 0_{N\times N}\\
0_{N\times N} & \sigma \dfrac{\delta t^{1/2}}{\sqrt{\delta x}} I_{N\times N}
\end{bmatrix}, \; \textrm{ and } \;
\underline{\omega}=
\begin{pmatrix}
\omega\\
\omega
\end{pmatrix},
\]
where $ I_{2N\times 2N}$ is the identity matrix of order $2N$ while $0_{N\times N}$ is an $N$th order null matrix. $0_{N}$ is an $N$th dimensional null vector. $F(t_{n})\in \mathbb{R}^{2N\times 2N}$, $G(t_{n})\in \mathbb{R}^{2N\times 2N}$ and $\underline{\omega}\in \mathbb{R}^{N}$. 

\begin{figure}[htp]
\centering
\subfigure[Contour plot for a single realisation of the stochastic wave equation][]{
\includegraphics[width=0.4\textwidth]{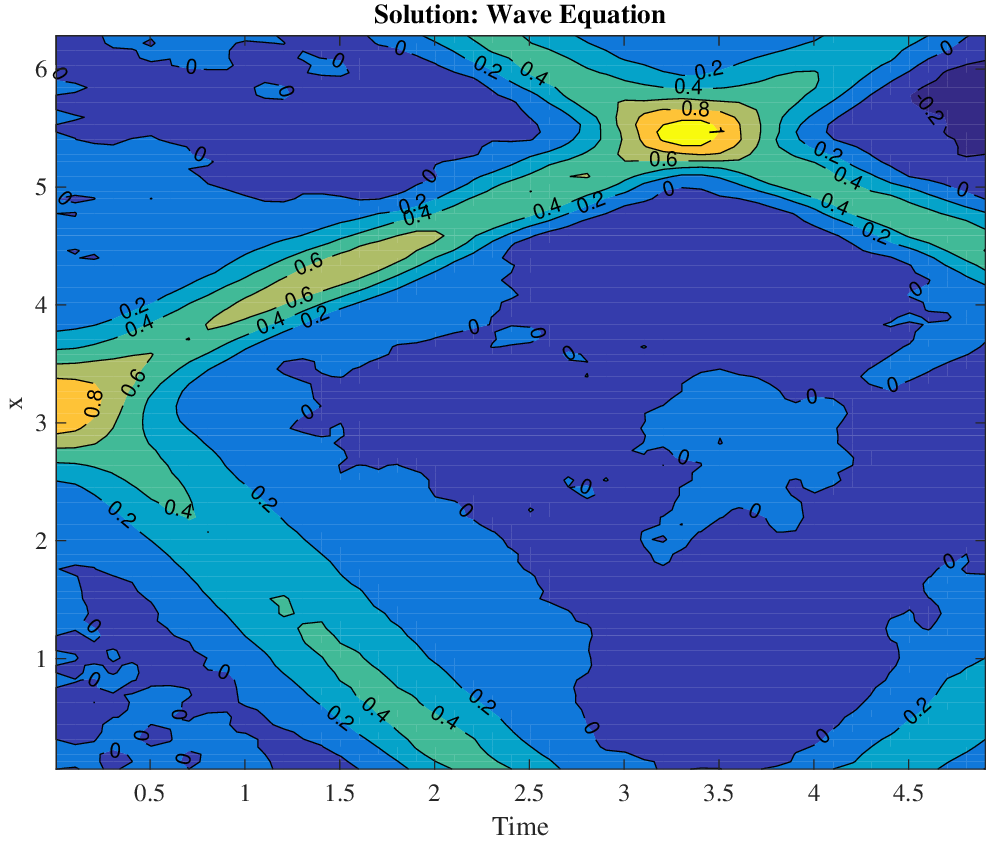}
\label{fig:subfig wavr1}}
\subfigure[Surface plot for a single realisation of the stochastic wave equation][ ]{
\includegraphics[width=0.45\textwidth]{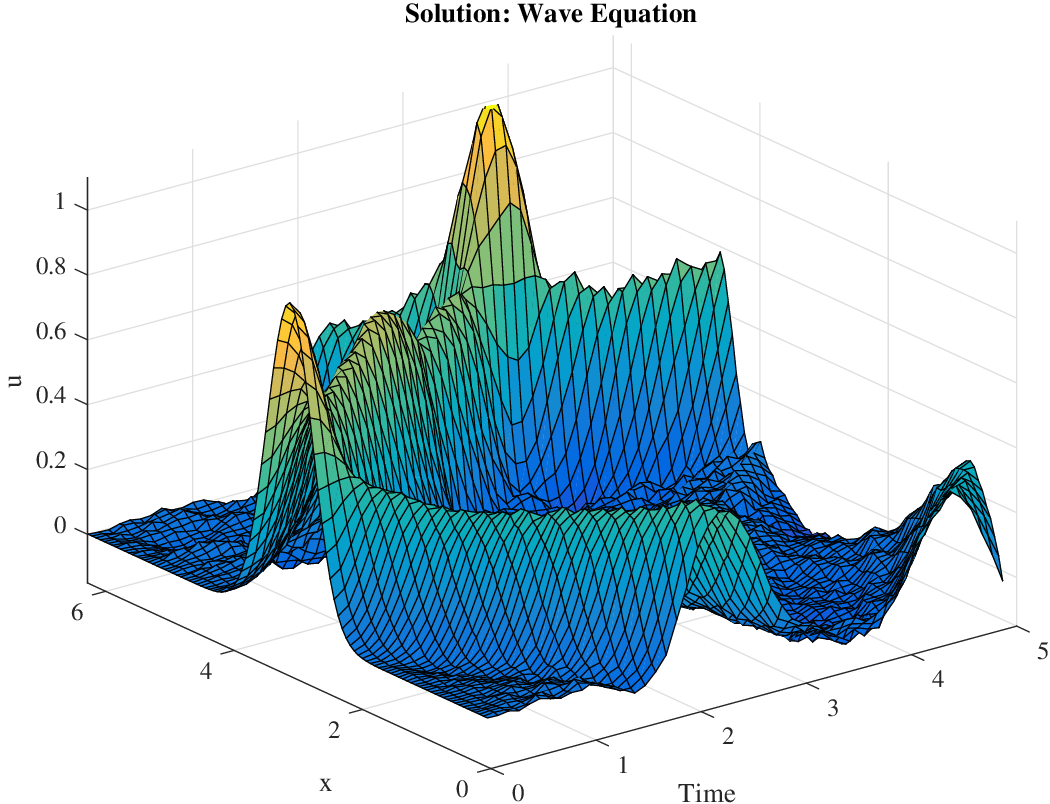}
\label{fig:subfig wavr2}}
\caption[Plots showing a single realisation of the stochastic wave equation]{Plots for a single realisation of the stochastic wave equation. The settings are: $L=2\pi$, $N=100$, $\delta x=L/N$, $\delta t=0.005$, and $T=1000$, $\mu=0.01$, $\sigma=0.2$, $C(x)=e^{\lambda(x)}$ where $\lambda(x)=\sin(x)$ and $u_{t_{0}}=\exp(-4(x-0.5L).^{2})$.}
\label{fig:globfig wavr7}
\end{figure}

The contour and surface plots for a single realisation of stochastic wave equation are in \Cref{fig:globfig wavr7}.

For the purpose of estimating the speed of the wave, which varies spatially, we give a little prelude to estimation of spatially varying parameters in the next section.

\section{Estimating spatially varying parameters }\label{Sec:7.2}
Varying parameters exist naturally, an example being in the speed of a wave moving in a heterogeneous media. Varying parameters present a challenge owing to their varying nature as opposed to static (in time) parameters. There are three broad categories of varying parameters: spatially varying parameters, parameters that vary with time and parameters that vary both with time and space. Estimation of time-varying parameters, albeit for deterministic models, and application to estimation of parameters in HIV/AIDS model, appears in \cite{Che}---in which least-squares methods have been used. Although consistent estimates are realized, extension to non-linear models remains to be realised. In this chapter, we study spatially varying parameters using filtering algorithms. We also provide applications in order to evaluate the performance of the algorithms introduced here.     

The strategy employed in this study comprises of three steps:
\begin{enumerate}
\item Express the parameter $\theta(x)$ as a Fourier series with a given number of modes, say, $N_{m}$ and collect all the coefficients in a vector $\lambda$.
\item By means of an appropriate filtering algorithm, estimate the vector of hyper-parameters, $\lambda$.
\item Substitute the estimated constant coefficients back in the Fourier series to obtain an estimate of the parameter $\theta(x)$. 
\end{enumerate}

To illustrate this method, we employ it to estimate the velocity of a wave travelling in a heterogeneous media. Let the parameter, $C(x)$, $x\in \mathbb{R}^{N}$, where $N$ is the number of dimensions, be given by
\begin{equation}
C(x) = \exp(f(x)),
\end{equation}
where
\begin{equation}
f(x)=a_{0}+\sum_{k=1}^{N_{m}}\dfrac{a_{k}}{k^{2}}\sin(k x)+\sum_{k=1}^{N_{m}}\dfrac{b_{k}}{k^{2}}\cos(k x), \; \; k=1, \; 2,\; 3, \; ..., \; N_{m},
\end{equation}
where the coefficients $a_{0}$, $a_{k}$ and $b_{k}$ are drawn randomly from a normal distribution of mean $0$ and variance $1$. This way, the parameter $C(x)$ will be positive for all values of $x$. The aim of this section is to obtain an estimate $C(x)_{est}$ of the wave velocity $C(x)$ by means of Kalman Bucy filter (KBF) and ensemble Kalman Bucy filter (EnKBF) \cite{Ang1}. To this end, we use a function 
\begin{equation}
g(x)=A_{0}+\sum_{k=1}^{\aleph_{m}}\dfrac{A_{k}}{k^{2}}\sin(k x)+\sum_{k=1}^{\aleph_{m}}\dfrac{B_{k}}{k^{2}}\cos(k x), \; \; k=1, \; 2,\; 3, \; ..., \; \aleph_{m},
\end{equation}
where $\aleph_{m} \leq N_{m}$ so that 
\begin{equation}
C(x)_{est}=\exp(g(x)).
\end{equation}

Let $\lambda\in \mathbb{R}^{2\aleph_{m}+1}$ be a vector whose elements are the coefficients of the function $g(x)$. That is,
\begin{equation}
\lambda=
\begin{pmatrix}
A_{0}\\
A_{1}\\
B_{1}\\
A_{2}\\
B_{2}\\
\vdotswithin{A_{\aleph_{modes}}}\\
A_{\aleph_{m}}\\
B_{\aleph_{m}}
\end{pmatrix}. \label{eq:6.1.6}
\end{equation}
Estimation of the spatially varying parameter, $C(x)$, is equivalent to estimating the parameters that form the vector $\lambda$. We consider two methods: using the likelihood with Metropolis-Hastings method and using a dual filter. Let us now consider each method in turn.

\section{Using the Likelihood with Metropolis-Hastings method } \label{subsec:5.4.1}

We now adapt \Cref{alg:05.1} to estimating the vector of parameters, $\lambda$. We let each parameter $\lambda_{i}$ have artificial dynamics with the transition density 
\begin{equation}
\zeta_{k+1}(\lambda_{i,k+1} \mid \lambda_{i,k}) = \mathcal{N}  \left( \lambda_{i,k}\cos(\phi), \, \dfrac{\omega}{\aleph} \sin(\phi) \right),
\end{equation}
where $\aleph$ is the mode number and the constants $\phi$ and $\omega$ are to be chosen. The density $\pi$ is defined by the filter likelihood, $$\pi(\delta y_{[t_{0},T]} \mid \tilde{u}_{[t_{0},T]}, \lambda_{k})= \underset{N \to \infty}{\underset{\delta t\to 0}{\textrm{l.i.m. }}} \pi_{t_{0}:t_{N}}(\delta y_{t_{0}:t_{N}} \mid \tilde{u}_{t_{0}:t_{N}}, \lambda_{k}),$$where $\tilde{u}_{t_{n}}$ is the filter prediction of the state at time $t_{n}$. Notice that the parameters, $\lambda$, are implicitly contained in the dynamics.

\begin{exam}[Advection Equation]{exa:example731}
\vspace{-0.4cm}
We take the advection equation, of \Cref{subsec:7.1.1}, and the given initial conditions. Furthermore, let there be time-continuous measurements of the state, $u$, given by,
\begin{equation}
dy_{t} = H(t)u(x,t)dt + Q(t)d\eta_{x,t}, \label{eq:7.2.4}
\end{equation}
where $\{\eta_{t,x}\}$ is standard space-time Brownian motion process. The initial value of $u_{t}$, $\{\beta_{t}\}$ and $\{\eta_{t,x}\}$ are uncorrelated. The aim is to estimate the spatially varying velocity, $C(x)$, by means of filter likelihood and Metropolis-Hastings algorithm.  
\end{exam}
We follow the discretisation described in \Cref{subsec:7.1.1} for the advection equation. 

 The measurements' equation, \cref{eq:7.2.4}, is discretised in time using Euler-Maruyama scheme to yield
  \begin{equation}
  \delta y_{t_{n}} = H(t_{n})u(x,t_{n})\delta t + R^{1/2}(t_{n})\delta \eta_{t_{n}}, \label{eq:7.2.5}
  \end{equation}
upon substituting $R= QQ^{\textrm{T}}/\delta x$ and where $\mathbb{E}[\delta \eta_{t_{n}}\delta \eta^{\textrm{T}}_{t_{n}}] = \textbf{I}_{r \times r}\delta t$. 
\begin{remark}
 The observation likelihood pdf for the KBF is Gaussian since the initial condition and the observation errors are Gaussian. So is the posterior pdf. With observation increments expressed as in \cref{eq:7.2.5}, the observation increment likelihood pdf is
 \begin{equation}
 \pi (\delta y_{t_{0}:t_{N}}| \tilde{u}_{t_{N}}, \lambda_{k}) \propto \prod_{n=0}^{N} \exp(-\dfrac{1}{2}\parallel\delta y_{t_{n}}- H(t_{n})\tilde{u}_{t_{n}}\delta t\parallel^{2}_{\delta t R(t_{n})}). \label{eq:7.3.4}
 \end{equation}
 Similarly, the filter forecast pdf is
\begin{equation}
 \pi( u_{t_{N}}|\delta y_{t_{N-1}:t_{0}}) \propto \prod_{n=0}^{N}\exp(-\dfrac{1}{2}\parallel u_{t_{n}}-\tilde{u}_{t_{n}}\parallel^{2}_{P_{t_{n}}}).
 \end{equation}
\end{remark}

The next step is to implement a KBF and EnKBF and to obtain the likelihood at each time steps. This is followed by implementing a Metropolis-Hastings algorithm. 

\begin{algorithm}[htp]
\caption{KBF likelihood with MH}
\label{alg:7.3.2}
\begin{algorithmic}[1]
\REQUIRE $\delta t$, $\aleph_{m}$, $N$, $u_{t_{0}}$, $\pi_{t_{n}}$ and $\lambda_{k}$.  
\ENSURE $\{\lambda_{k}\}_{k=1}^{T}$.
\FOR{$k=1$ \TO $N$}
\STATE{Draw $\tilde{\lambda} \sim \mathcal{N}  \left( \lambda_{i,k}\cos(\phi), \, \dfrac{\omega}{\aleph_{m}} \sin(\phi) \right) \quad \forall i=1,\; 2, \; ...,\; 2\aleph_{m} +1 $.}
\STATE{Compute $C_{k}(x)=\exp(g(x, \lambda_{k}))$.}
\FOR{$n=1$ \TO $T$, $\delta t>0$}
\STATE{Run a single step KBF prediction mean $\tilde{u}_{t_{n+1}}  =u_{t_{n}} + F(t_{n},\lambda_{k}) u_{t_{n}}\delta t$}
\STATE{Run a single step KBF prediction covariance $\tilde{P}_{t_{n+1}} = P_{t_{n}} + F(t_{n},\lambda_{k}) P_{t_{n}}\delta t+ P_{t_{n}}F^{\textrm{T}}(t_{n},\lambda_{k}) \delta t + G(t_{n})G^{\textrm{T}}(t_{n}) \delta t$}
\STATE{Run a single step KBF analysis mean $u_{t_{n+1}}  = \tilde{u}_{t_{n+1}} + P_{t_{n}}H^{\textrm{T}}(t_{n}) R^{-1}(t_{n})(dy_{t_{n}}-H(t_{n}, \lambda_{k})\tilde{u}_{t_{n}}\delta t)$}
\STATE{Run a single step KBF analysis covariance $P_{t_{n+1}} =\tilde{P}_{t_{n+1}}  + G(t_{n})G^{\textrm{T}}(t_{n}) \delta t- \tilde{P}_{t_{n}}H^{\textrm{T}}(t_{n}) R^{-1}(t_{n})H(t_{n})P_{t_{n}}\delta t $}
\ENDFOR
\STATE{\underline{Metropolis Hastings}}
\STATE{Compute $\alpha_{\textrm{ratio}}=\dfrac{\pi (\delta y_{t_{N}:t_{0}}| \tilde{u}_{t_{N}}, \lambda_{k})}{\pi (\delta y_{t_{N}:t_{0}}| \tilde{u}_{t_{N}}, \lambda_{k-1})}\dfrac{\rho_{\lambda}(\tilde{\lambda})}{\rho_{\lambda}(\lambda_{k})}$}
\STATE{Compute $\alpha=\min(1, \alpha_{\textrm{ratio}})$}
\IF{$\alpha>U(0,1)$}
\STATE{$\tilde{\lambda}=\lambda_{k}$}
\ELSE
\STATE{$\lambda_{k}=\tilde{\lambda}$}
\ENDIF
\ENDFOR
\end{algorithmic} 
\end{algorithm}

The same is repeated but with EnKBF in the place of KBF. \Cref{alg:7.3.3} is pseudo code showing the basic steps.

\begin{algorithm}[htp]
\caption{EnKBF likelihood with MH}
\label{alg:7.3.3}
\begin{algorithmic}[1]
\REQUIRE $\delta t$, $\aleph_{m}$, $M$, $N$, $\{u^{i}_{t_{0}}\}_{i=1}^{M}$, $\pi_{t_{n}}$ and $\lambda_{0}$.  
\ENSURE $\{\lambda_{k}\}_{k=1}^{T}$.
\FOR{$k=1$ \TO $N$}
\STATE{Draw $\tilde{\lambda} \sim \mathcal{N}  \left( \lambda_{i,k}\cos(\phi), \, \dfrac{\omega}{\aleph_{m}} \sin(\phi) \right) \quad \forall i=1,\; 2, \; ...,\; 2\aleph_{m} +1 $.}
\STATE{Compute $C_{k}(x)=\exp(g(x, \lambda_{k}))$.}
\FOR{$n=1$ \TO $T$, $\delta t>0$}
\FOR{$i=1$ \TO $M$}
\STATE{Run a single step EnKBF prediction ensemble $\tilde{u}^{i}_{t_{n+1}}  =u^{i}_{t_{n}} + F(t_{n}, \lambda_{k})u^{i}_{t_{n}}\delta t$}
\STATE{Run a single step EnKBF analysis ensemble $u^{i}_{t_{n+1}}  = \tilde{u}^{i}_{t_{n+1}} + P_{t_{n}}H^{\textrm{T}}(t_{n}) R^{-1}(t_{n})(dy_{t_{n}} + \epsilon _{i} -H(t_{n})\tilde{u}^{i}_{t_{n}}\delta t)$}
\ENDFOR
\STATE{Compute prediction ensemble mean: $\bar{u}_{t_{n}} = \dfrac{1}{M} \sum_{i=1}^{M} \tilde{u}^{i}_{t_{n}}$.}
\STATE{Compute analysis ensemble mean: $\hat{u}_{t_{n}} = \dfrac{1}{M} \sum_{i=1}^{M} u^{i}_{t_{n}}$.}
\STATE{Compute covariance: $P_{t_{n}} = \dfrac{1}{M-1} \sum_{i=1}^{M} (u^{i}_{t_{n}}-\hat{u}_{t_{n}})(u^{i}_{t_{n}}-\hat{u}_{t_{n}})^{\textrm{T}}$.}
\ENDFOR
\STATE{\underline{Metropolis Hastings}}
\STATE{Compute $\alpha_{\textrm{ratio}}=\dfrac{\pi (\delta y_{t_{N}:t_{0}}| \bar{u}_{t_{N}},\lambda_{k})}{\pi (\delta y_{t_{N}:t_{0}}| \bar{u}_{t_{N}}, \lambda_{k-1})}\dfrac{\rho_{\lambda}(\tilde{\lambda})}{\rho_{\lambda}(\lambda_{k})}$}
\STATE{Compute $\alpha=\min(1, \alpha_{\textrm{ratio}})$}
\IF{$\alpha>U(0,1)$}
\STATE{$\tilde{\lambda}=\lambda_{k}$}
\ELSE
\STATE{$\lambda_{k}=\tilde{\lambda}$}
\ENDIF
\ENDFOR
\end{algorithmic} 
\end{algorithm}

\begin{figure}[htp]
\centering
\subfigure[Estimate of parameter $A_{0}$][]{
\includegraphics[width=0.45\textwidth]{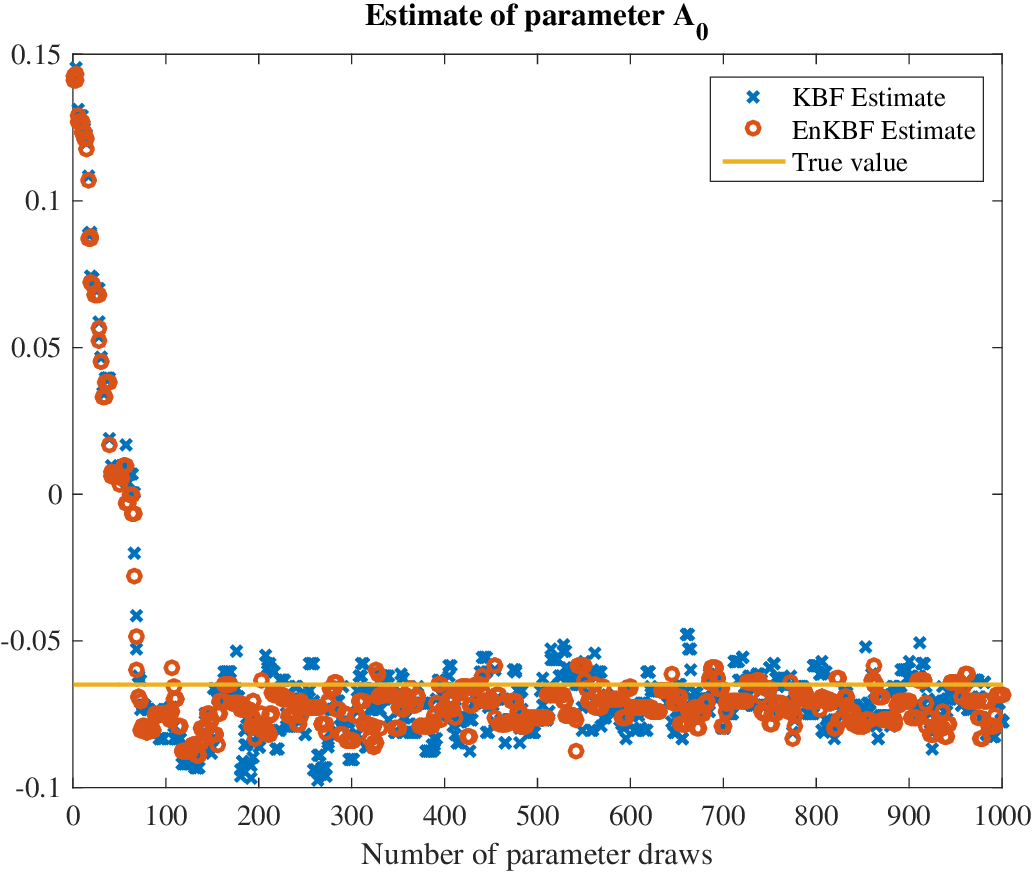}
\label{fig:subfig dr1}}
\subfigure[Estimate of parameter $A_{1}$][]{
\includegraphics[width=0.45\textwidth]{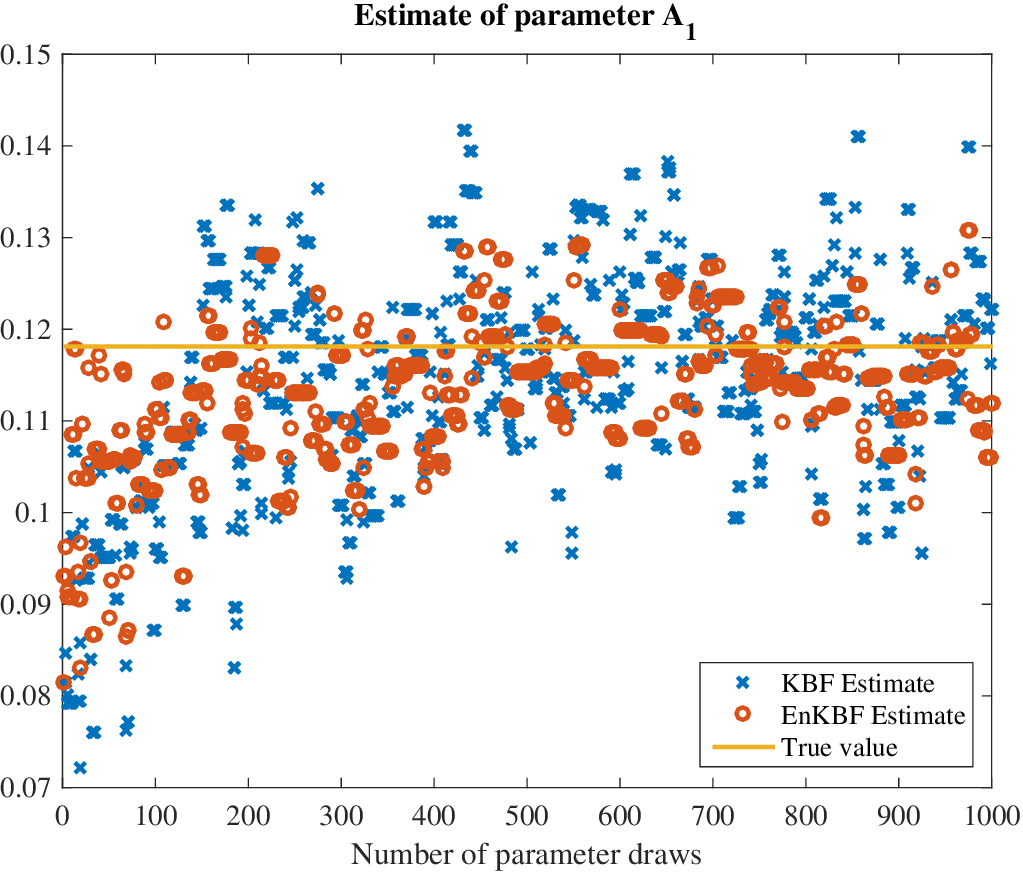}
\label{fig:subfig dr4}}
\caption[Plots for velocity parameters for the first two parameters in $\lambda$ obtained using KBF  and EnKBF]{Plots for velocity parameters for the first three parameters in $\lambda$ obtained using KBF  and EnKBF and  run for $1000$ time steps and for $1000$ Metropolis-Hastings cycles. The number of particles used for EnKBF is $M=1000$, the time step size used in both filters is $dt=0.01$, $\mu=0.001$ and $100$ grid points are used. The plots indicate that the estimates, for both filters, converge after about $100$ iterations.}
\label{fig:globfig dar1}
\end{figure}

Results for the first $2$ parameters are shown in \Cref{fig:globfig dar1}. The results in \Cref{fig:globfig dar1} show that the EnKBF performs like the KBF filter. This agrees with the theory (see, for example, \cite{Ang1}), that the EnKBF yields optimal results in the limit $M \to\infty$. It is also noteworthy that Metropolis-Hastings algorithm converges to the true parameter estimate. This can be seen in \Cref{fig:subfig dr1}, for example, where the filter estimate converges after about $100$ parameter draws.

We now look at the errors in the parameter estimates, the better to see the performance of the filters for the $21$ hyper-parameters.

\begin{figure}[htp]
\centering
\subfigure[Parameter estimation error: Metropolis Hastings][]{
\includegraphics[width=0.45\textwidth]{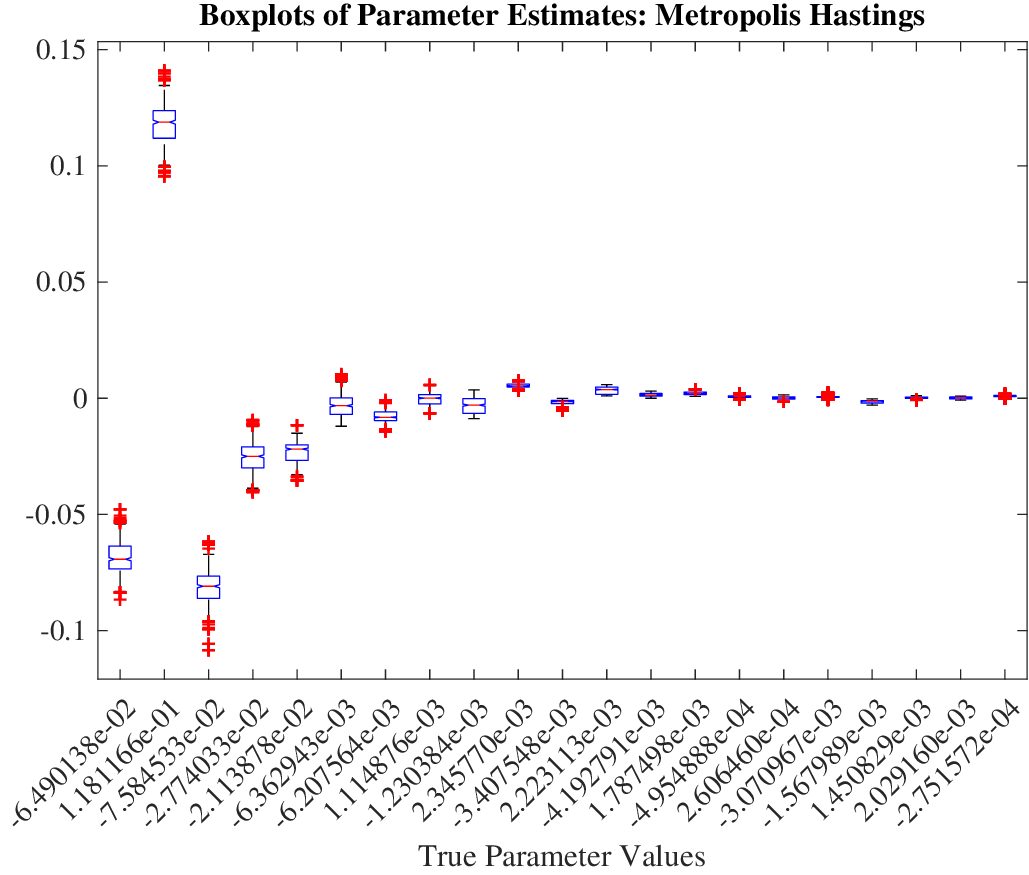}
\label{fig:subfig dr12}}
\subfigure[Box plots of parameter estimation: Metropolis Hastings][]{
\includegraphics[width=0.45\textwidth]{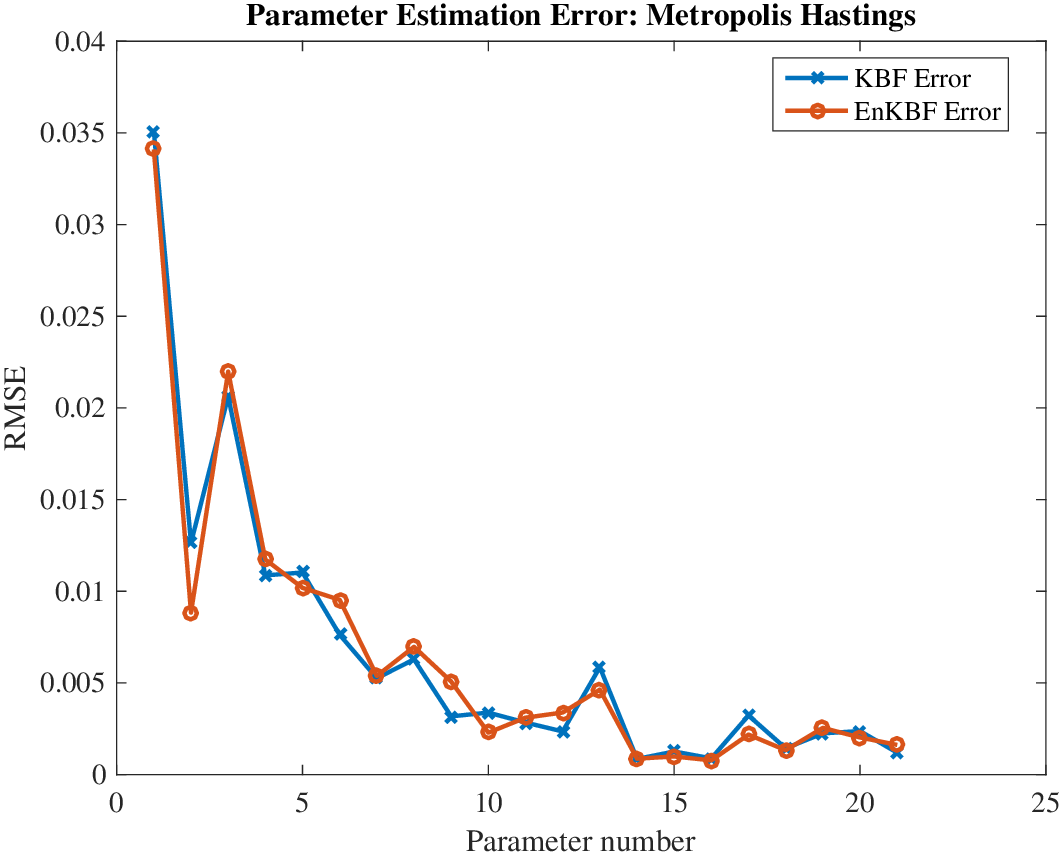}
\label{fig:subfig dr11}}
\caption[A box plot of the $21$ hyper-parameters estimated using EnKBF and a plot of the root mean square error for the parameter estimates obtained using EnKBF and KBF, respectively: advection equation]{(a) Box plot for the $21$ velocity parameters for the EnKBF run for $1000$ time steps with $1000$ Metropolis-Hastings cycles. A burn-in of $500$ parameter draws is discarded. The stochastic advection equation model is used with the following settings: $L=2\pi$, $100$ grid points, $\delta t=0.01$, $M=1000$ particles, $\mu = 0.001$ and localization radius of $10$ grid points. (b) A plot of the root mean square error for the $21$ hyper-parameter estimates obtained using EnKBF and KBF, respectively. The plot indicates that the performance of EnKBF matches that of KBF in this setting.   }
\label{fig:globfig dar11}
\end{figure}

In \Cref{fig:subfig dr11} panels are plotted the box-plots showing the dispersion of parameter estimates resulting from the use of EnKBF and the root mean square errors for parameter estimates for both the EnKBF and KBF. The RMSE values are computed as follows.
\begin{equation}
\textrm{RMSE} = \sqrt{\dfrac{1}{N}\sum_{k=1}^{N} (\lambda_{i,k} - \lambda^{\textrm{true}}_{i})^{2}},
\end{equation}
where $\lambda_{i,k}$ is the estimate of the $i$th parameter at Metropolis-Hastings cycle $k$ and $\lambda^{\textrm{true}}_{i}$ the true $i$th parameter.

The RMSE for both the KBF and EnKBF, as shown in \Cref{fig:subfig dr11}, indicate that the performance of EnKBF matches that of the KBF for the $21$ parameters. These heuristic results corroborate the theoretical findings. The boxplot, \Cref{fig:subfig dr12}, shows the dispersion of parameter samples in the case when EnKBF is used. The result indicates that the estimates matches the true parameter values, with not so many outliers. This is indicative not only of the performance of EnKBF but also the Metropolis-Hastings procedure in locating the true parameter values and ensuring that no large excursions are made from the true parameter values.

We now implement \Cref{alg:7.3.2,alg:7.3.3} for the discretised wave equation, \cref{eq:7.1.13}.

\begin{exam}[Wave Equation]{exam:6}
\vspace{-0.4cm}
We take the wave equation of \Cref{subsec:7.3.2} and the associated initial conditions, \cref{eq:7.1.11a,eq:7.1.11b}. The measurements are given by \cref{eq:7.2.4}. The initial value of $u_{t}$, $\{\beta_{t}\}$ and $\{\eta_{t}\}$ are uncorrelated. The aim is to estimate the spatially varying velocity, $C(x)$, by means of filter likelihood and Metropolis-Hastings algorithm.  
\end{exam}

\begin{figure}[htp]
\centering
\subfigure[Estimate of parameter $A_{0}$][]{
\includegraphics[width=0.45\textwidth]{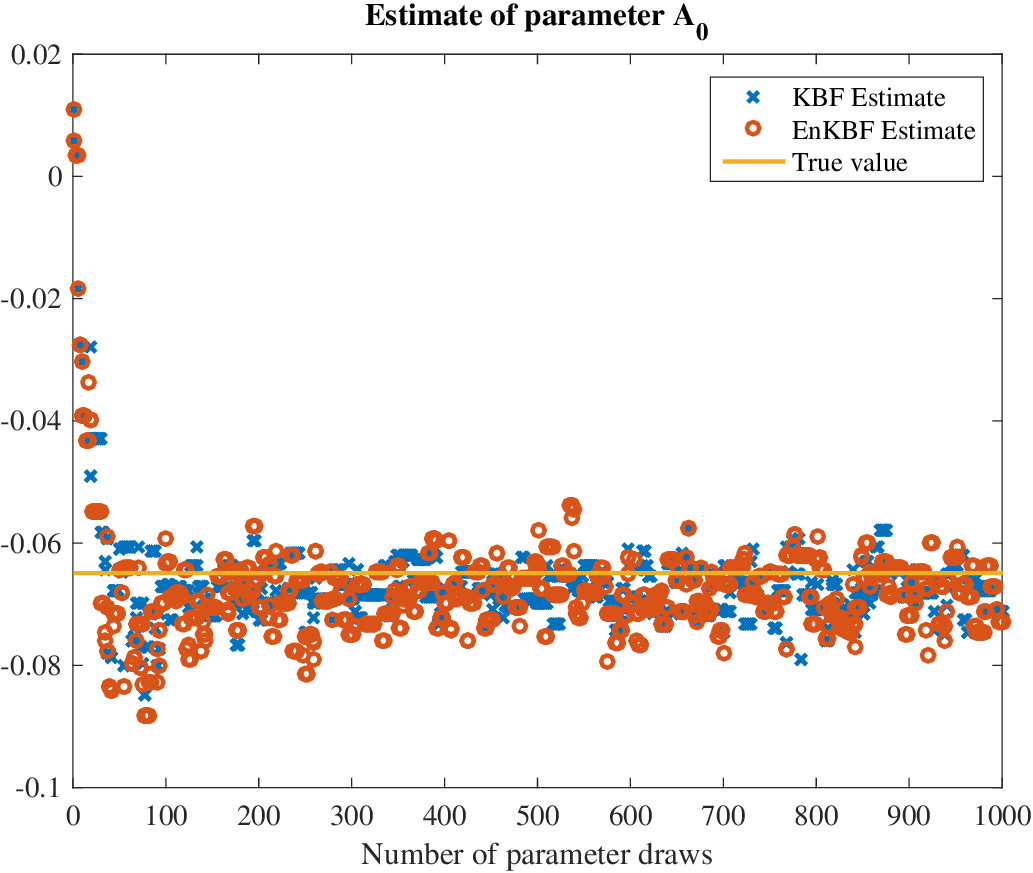}
\label{fig:subfig drW1}}
\subfigure[Estimate of parameter $A_{1}$][]{
\includegraphics[width=0.45\textwidth]{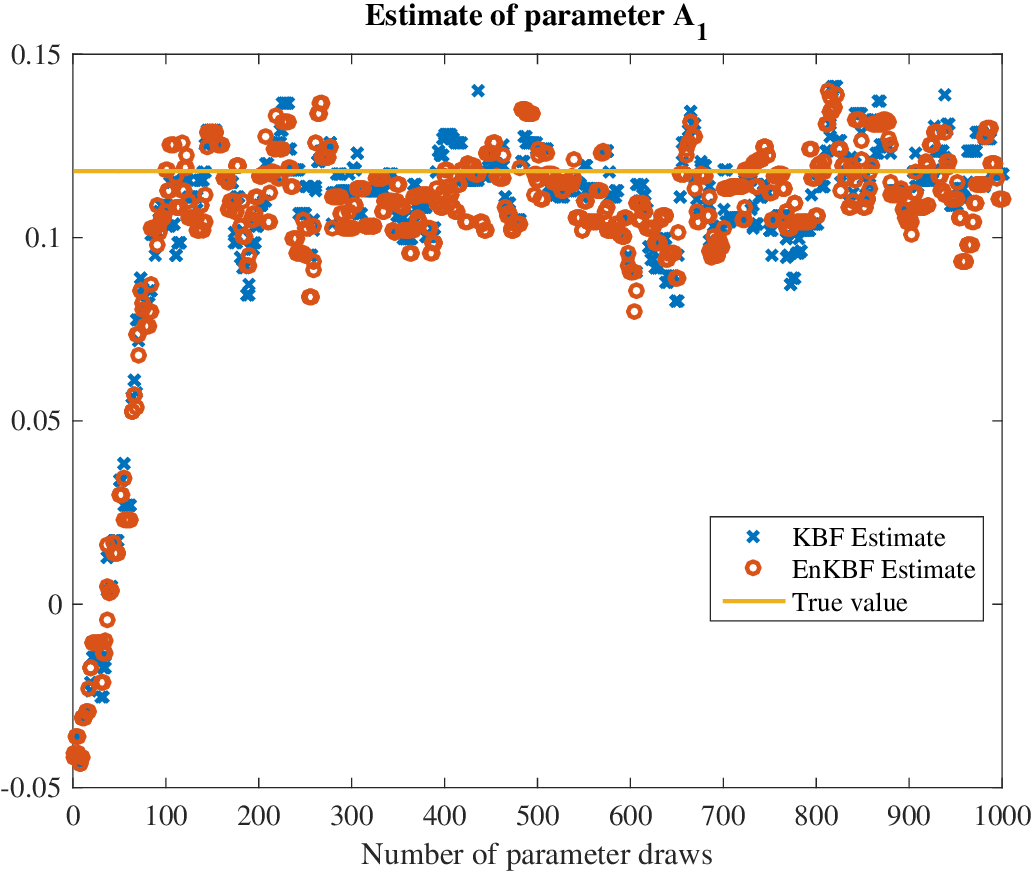}
\label{fig:subfig drW4}}
\caption[Plots for velocity parameters for the first two parameters in $\lambda$ obtained using KBF  and EnKBF.]{Plots for velocity parameters for the first three parameters in $\lambda$ obtained using KBF  and EnKBF and  run for $1000$ time steps and $1000$ Metropolis-Hastings cycles. The number of particles used for EnKBF is $M=1000$, the time step size used in both filters is $dt=0.01$, $\mu=0.001$ and $100$ grid points are used. The plots indicate that the estimates, for both filters, converge after about $100$ iterations.}
\label{fig:globfig dar1W}
\end{figure}

The discretisation of the wave equation is as shown in \Cref{subsec:7.3.2}. The panels in \Cref{fig:globfig dar1W} show the results.

The results in \Cref{fig:globfig dar1W} indicate a close match in the performance of the EnKBF and KBF. This is as was anticipated in the theoretical findings, some of which are found in \cite{Ang1}. Notice also the the two filters converge to the true parameter values after a few parameter draws (about $50$ in \Cref{fig:subfig drW1} and $100$ in \Cref{fig:subfig drW4})---which is indicative of the robustness of the Metropolis-Hastings algorithm atop the EnKBF and KBF filters. The results also show that there are no wide excursions from the true parameter values, which testifies of the good performance of \Cref{alg:7.3.2,alg:7.3.3}.

\begin{figure}[htp]
\centering
\subfigure[Boxplots of parameter estimation: Metropolis Hastings][]{
\includegraphics[width=0.45\textwidth]{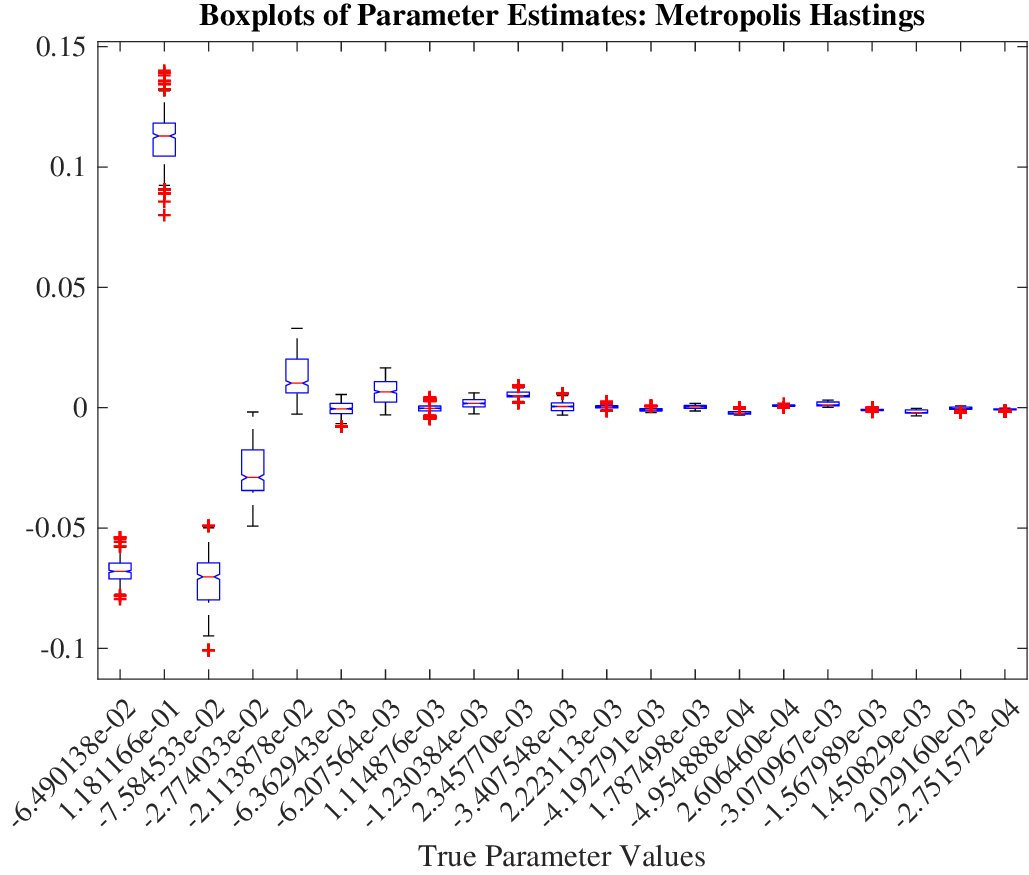}
\label{fig:subfig drW12}}
\subfigure[Parameter estimation error: Metropolis Hastings][]{
\includegraphics[width=0.45\textwidth]{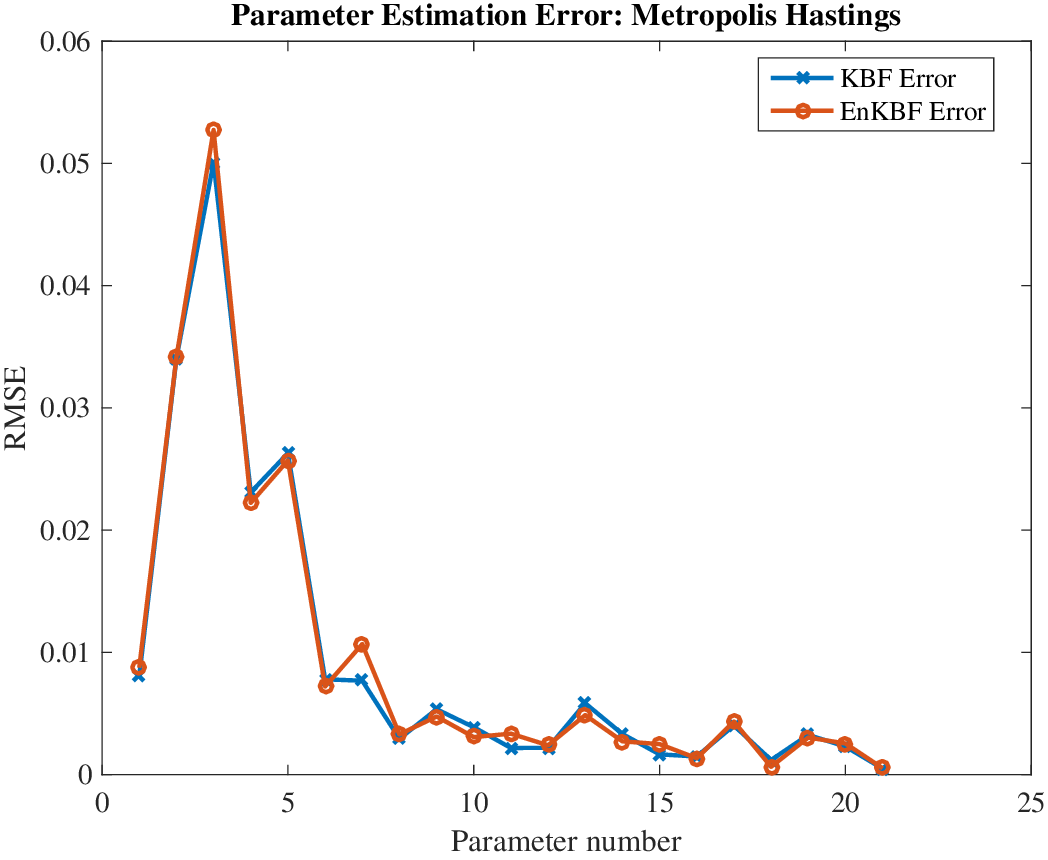}
\label{fig:subfig drW11}}
\caption[A box plot of the $21$ hyper-parameters and a plot of the root mean square error for the parameter estimates obtained using EnKBF and KBF, respectively: wave equation]{(a) Box plot for velocity parameters for the EnKBF run for $1000$ time steps and $1000$ Metropolis-Hastings cycles. A burn-in of $500$ parameter draws is discarded. The stochastic wave equation model is used with the following settings: $L=2\pi$, $100$ grid-points, $\delta t=0.01$, $M=1000$ particles, $\mu = 0.001$ and localization radius of $10$ grid points. (b) A plot of the root mean square error for the parameter estimates obtained using EnKBF and KBF, respectively. The plot indicates that the performance of EnKBF matches that of KBF in this setting.}
\label{fig:globfig dar11W}
\end{figure}

In \Cref{fig:globfig dar11W} are plotted the box-plots showing the dispersion of the $21$ hyper-parameter estimates resulting from the use of EnKBF and the root mean square errors for parameter estimates for both the EnKBF and KBF. 

The EnKBF elicits an optimal performance as can be seen in \Cref{fig:subfig drW11} where the RMSEs for both the EnKBF and the optimal filter (KBF) match for all the $21$ hyper-parameters. This also, as in the advection equation above, is in agreement with the theoretical findings that the EnKBF attains an optimal estimate in the limit $M\to \infty$. The boxplot, \Cref{fig:subfig drW12}, shows that the mean of EnKBF parameter estimates matches the true parameter values. What is more, there are not many outliers in the estimates. All this show that the EnKBF-Metropolis-Hastings algorithm is robust. 

In the next section, we apply the concepts on dual estimation to stochastic hyperbolic PDEs, which, in this case, are the advection and the wave equations.

\section{Simultaneous estimation of the state and spatially varying parameters}
The dual filter (see \Cref{subsec:5.3.2} for details) can be adapted to allow for estimating both the state and spatially varying parameters, contemporaneously. The idea here is to replace the parameters in the dual filter with hyper-parameters of the varying parameters to be approximated. The hyper-parameters are then updated simultaneously, and in parallel, with the state at each iteration, where one filter estimates the state and the other filter updates the hyper-parameters---with each filters making use of the outcome of the other. To illustrate this argument, we turn to the advection and wave equation described in Examples 7.1 and 7.2, respectively, and use the KBF-EnKBF dual filter---in which the state is propagated and updated by means of the KBF whilst the hyper-parameters are updated using EnKBF---and ENKBF dual filter---where both the state and the hyper-parameters are propagated and updated using two EnKBFs running in parallel. The spatially varying velocity is as shown in \Cref{Sec:7.2}.

In the KBF-EnKBF dual filter, we update, for every $j$th parameter particle $\lambda^{j}_{t} \in \{\lambda_{t}^{j}\}_{j=1}^{L}$ the state estimate, $\hat{u}^{j}_{t}$, using the KBF; that is,
\begin{subequations}
\begin{align}
d\hat{u}^{j}_{t}  &= F(t)\hat{u}^{j}_{t}dt  + P_{t}H^{\textrm{T}}(t) R^{-1}(t)(dy_{t}-H(t)\hat{u}^{j}_{t}dt), \label{eq:7.1s.3a}\\
dP_{t} &= F(t)P_{t}dt+ P_{t}F^{\textrm{T}}(t)dt + G(t)G^{\textrm{T}}(t) dt- P_{t}H^{\textrm{T}}(t) R^{-1}(t)H(t)P_{t}dt.\label{eq:7.1s.3b}
\end{align} 
\end{subequations}
The parameters are updated using the EnKBF; that is, each parameter hypothesis, $\lambda^{j}_{t}$, is updated using
\begin{equation}
d\lambda^{j}_{t} = D^{L}_{t}H(t)R^{-1}(t)(dy_{t}-0.5(H(t)\hat{x}^{j}_{t} + H(t)\hat{x}_{t})dt); \quad t_{0} \leq t,\label{eq:7.3s.5}
\end{equation}
where 
\begin{subequations}
\begin{align}
\hat{\lambda}_{t} &= \dfrac{1}{L}\sum^{L}_{j=1}\lambda^{j}_{t} ; \qquad  t_{0} \leq t, \\
D^{L}_{t} &= \dfrac{1}{L-1}\sum^{L}_{i=1}(\lambda^{j}_{t}-\hat{\lambda}_{t})(\hat{u}^{j}_{t}- \hat{u}_{t})^{\textrm{T}} ; \qquad  t_{0} \leq t, \label{eq:7.3s.6a}
\intertext{where}
\hat{u}^{j}_{t} & = \dfrac{1}{M}\sum_{i=1}^{M}u^{i,j}_{t}, \label{eq:7.3s.6b} \\
\hat{u}_{t} & = \dfrac{1}{L}\sum_{i=1}^{L}\hat{u}^{j}_{t}.
\end{align}
\end{subequations}
\Cref{alg:07.4.2} gives a summary of the KBF-EnKBF dual filter.
\begin{algorithm}[htp]
\caption{KBF-EnKBF dual filter}
\label{alg:07.4.2}
\begin{algorithmic}[1]
\REQUIRE $u^{j}_{t_{0}}$, $\lambda^{j}_{t_{0}}$, $w^{j}_{t_{0}}= 1/L$\; $\forall  j \in \{1, \; 2, \; ..., \; L\}$, $P_{t_{0}}$ and $\delta y_{[t_{0},t_{T}]}$.  
\ENSURE $\hat{u}_{[t_{0},t_{N}]}$, $\hat{\lambda}_{[t_{0},t_{N}]}$.
\FOR{$n=1$ \TO $N$, $\delta t>0$} 
\FOR{$j=1$ \TO $L$} 
\STATE{Update $\hat{u}^{j}_{t_{n}}$ using \cref{eq:7.1s.3a,eq:7.1s.3b} }
\STATE{Update parameters $\lambda^{j}_{t_{n}}$ using \cref{eq:7.3s.5}}
\ENDFOR
\STATE{Compute $\hat{\lambda}_{t_{n}} = \dfrac{1}{L} \sum_{j=1}^{L}\lambda^{j}_{t_{n}}$ }
\STATE{Compute $\hat{u}_{t_{n}} = \dfrac{1}{L} \sum_{i=1}^{L} \hat{u}^{j}_{t_{n}}$ }
\ENDFOR
\end{algorithmic} 
\end{algorithm}

The EnKBF dual filter consists of an update of $M$ particles of the state, $u^{i,j}_{t} \in \{u^{i,j}_{t}\}_{i,j=1}^{M,L}$, for every parameter particle, $\lambda^{j}_{t} \in \{\lambda_{t}^{j}\}_{j=1}^{L}$, using the EnKBF; that is,
\begin{equation}
du^{i,j}_{t}  = F(t)u^{i,j}_{t}dt  +G(t)d\beta^{i,j}_{t} +   P^{M}_{t}H^{\textrm{T}}(t) R^{-1}(t)(dy_{t}+ R^{1/2}(t)\eta_{t}^{i,j}-H(t)u^{i,j}_{t}dt), \label{eq:7.2s.1}
\end{equation}
where $\{\eta^{i,j}_{t}, \, t_{0}\leq t\}$ and $\{\beta^{i,j}_{t}, \, t_{0}\leq t\}$ are, respectively, standard Brownian motion vector processes. The parameters are updated using the EnKBF given by \cref{eq:7.3s.5}. The summary of the EnKBF dual filter is given in \Cref{alg:07.4.1}.
\begin{algorithm}[htp]
\caption{EnKBF dual filter}
\label{alg:07.4.1}
\begin{algorithmic}[1]
\REQUIRE $u^{i,j}_{t_{0}}$, $\lambda^{j}_{t_{0}}$, $w^{j}_{t_{0}}= 1/L$\; $\forall i \in \{1, \; 2, \; ..., \; M\} \; j \in \{1, \; 2, \; ..., \; L\}$, $P_{t_{0}}$ and $\delta y_{[t_{0},t_{T}]}$.  
\ENSURE $\hat{u}_{[t_{0},t_{N}]}$, $\hat{\lambda}_{[t_{0},t_{N}]}$.
\FOR{$n=1$ \TO $N$, $\delta t>0$} 
\FOR{$j=1$ \TO $L$} 
\FOR{$i=1$ \TO $M$} 
\STATE{Calculate $u^{i,j}_{t_{n}}$ using \cref{eq:7.2s.1}  }
\ENDFOR
\STATE{Update $\hat{u}^{j}_{t_{n}}$ using \cref{eq:7.3s.6b} }
\STATE{Update parameters $\lambda^{j}_{t_{n}}$ using \cref{eq:7.3s.5}}
\ENDFOR
\STATE{Compute $\hat{\lambda}_{t_{n}} = \dfrac{1}{L} \sum_{j=1}^{L}\lambda^{j}_{t_{n}}$ }
\STATE{Compute $\hat{u}_{t_{n}} = \dfrac{1}{L} \sum_{i=1}^{L} \hat{u}^{j}_{t_{n}}$ }
\ENDFOR
\end{algorithmic} 
\end{algorithm}

\begin{figure}[htp]
\centering
\subfigure[Estimate of parameter $A_{0}$][]{
\includegraphics[width=0.45\textwidth]{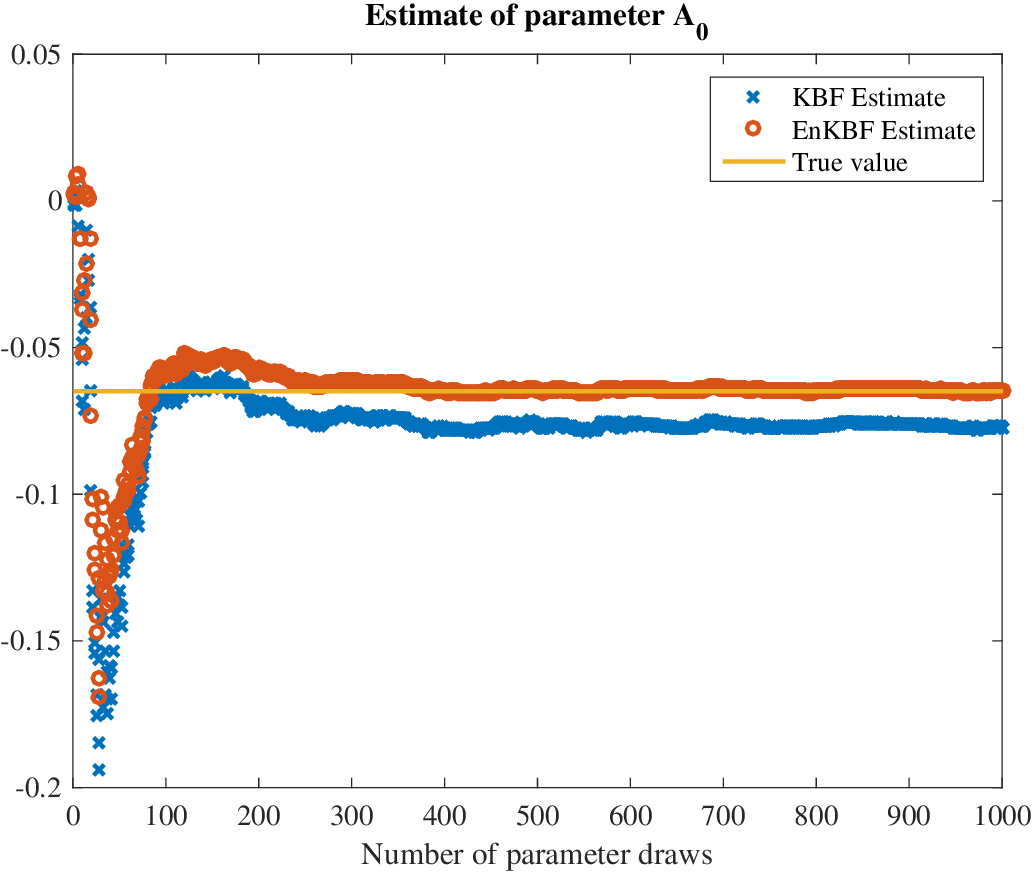}
\label{fig:subfig 7.7a}}
\subfigure[Estimate of parameter $A_{1}$][]{
\includegraphics[width=0.45\textwidth]{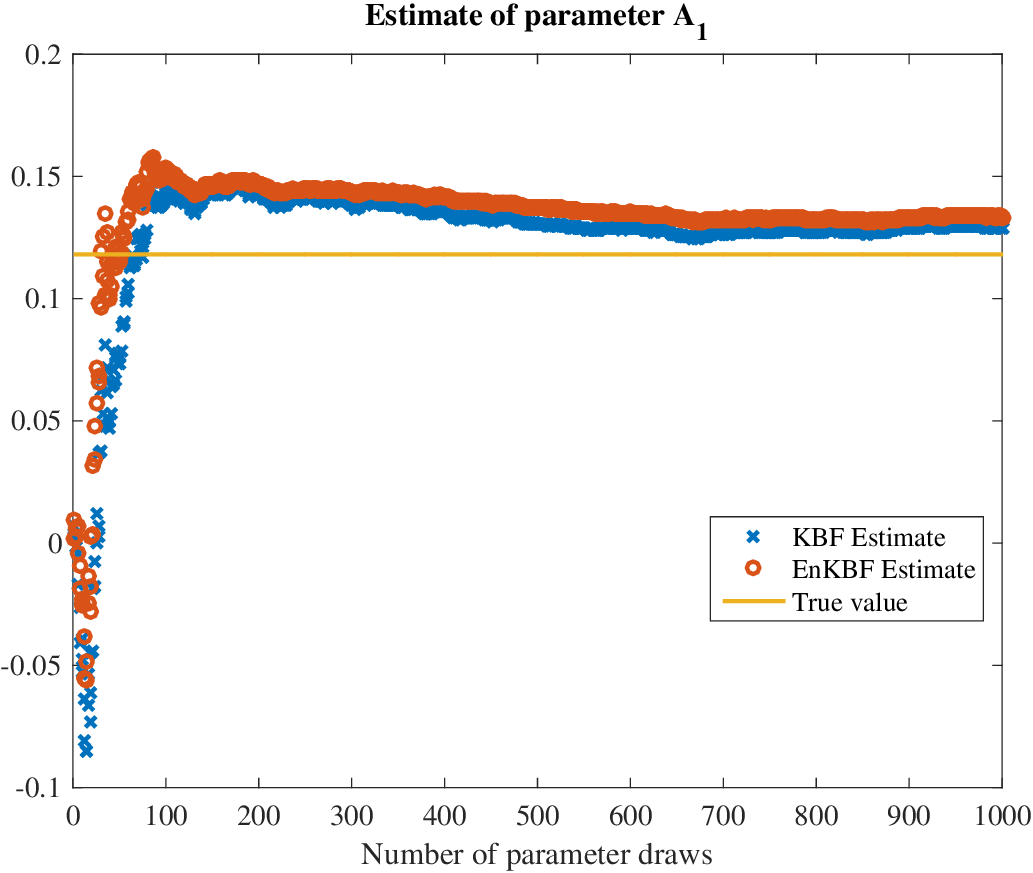}
\label{fig:subfig 7.7b}}
\caption[Plots for velocity parameters in the advection equation for the first two parameters in $\lambda$ obtained using KBF-EnKBF  and EnKBF dual filters, both run for $1000$ time steps]{(a), (b) and (c) are plots for velocity parameters for the first three parameters in $\lambda$ obtained using KBF-EnKBF  and EnKBF dual filters, applied to the advection equation, both run for $1000$ time steps. The number of particles, for the state and hyper-parameters, used in EnKBF is $M=1000$ and $L=1000$; the time step size used in both filters is $dt=0.01$. $\mu=0.001$ and $100$ grid points are used. Compared to the results obtained using Metropolis-Hastings algorithm (\Cref{subsec:5.4.1}), the dual filters register a better performance, at least in this example.}
\label{fig:globfig 7.7}
\end{figure}

The panels in \Cref{fig:globfig 7.7} show the results for the first two parameters in \cref{eq:6.1.6} when the dual filters are applied to the advection equation.

Evidently, from \Cref{fig:globfig 7.7}, the performance of the EnKBF and KBF-EnKBF dual filters almost match. Both filters converge to the true estimate after a few iterations (about $100$ in \Cref{fig:subfig 7.7a}). This is another testament to the fact that EnKBF attains optimal estimates at high ensemble values. Moreover, both the EnKBF and KBF-EnKBF parameter estimates are not much spread as compared to the previous case where Metropolis-Hastings was used. This is more evident in the following results for the $21$ hyper-parameters estimated.

In the following panels are plotted the root mean square errors for parameter estimates for both the EnKBF and KBF-EnKBF dual filters and the box-plots showing the dispersion of parameter estimates resulting from the use of EnKBF dual filter applied to the advection equation. 
\begin{figure}[htp]
\centering
\subfigure[Boxplots of parameter estimates: dual filters][]{
\includegraphics[width=0.45\textwidth]{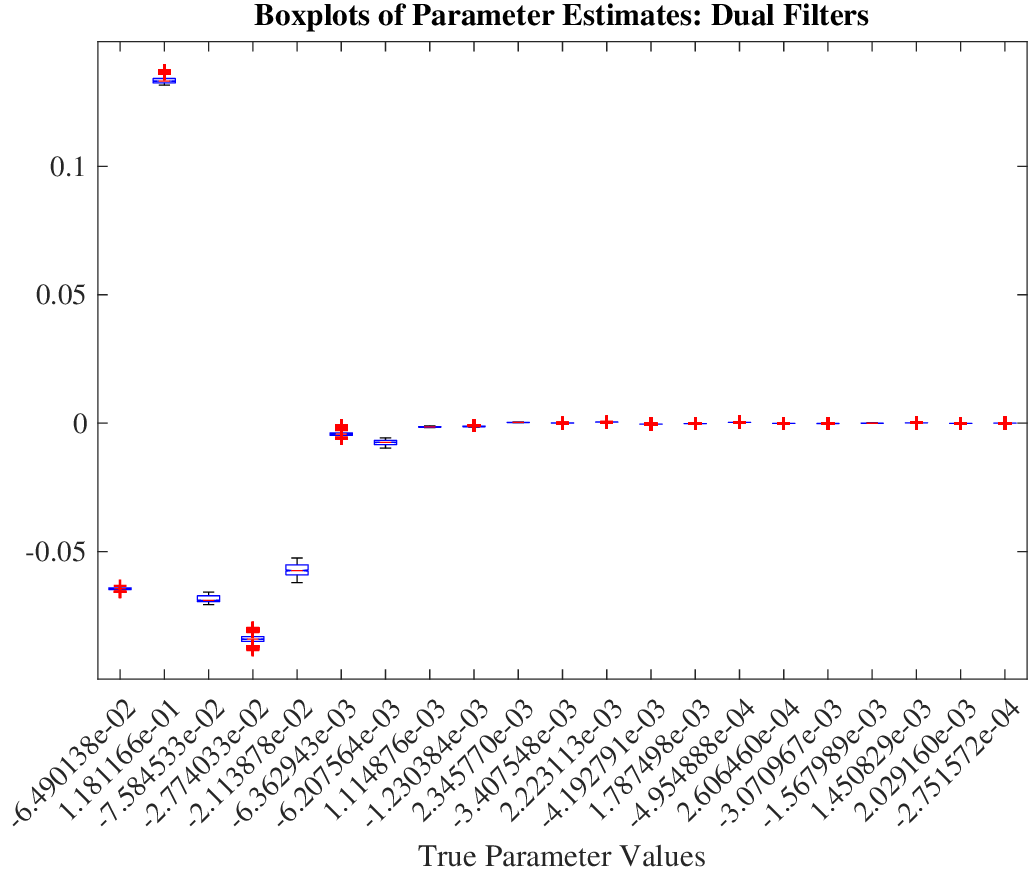}
\label{fig:subfig 7.8b}}
\subfigure[Parameter estimation error: dual filters][]{
\includegraphics[width=0.45\textwidth]{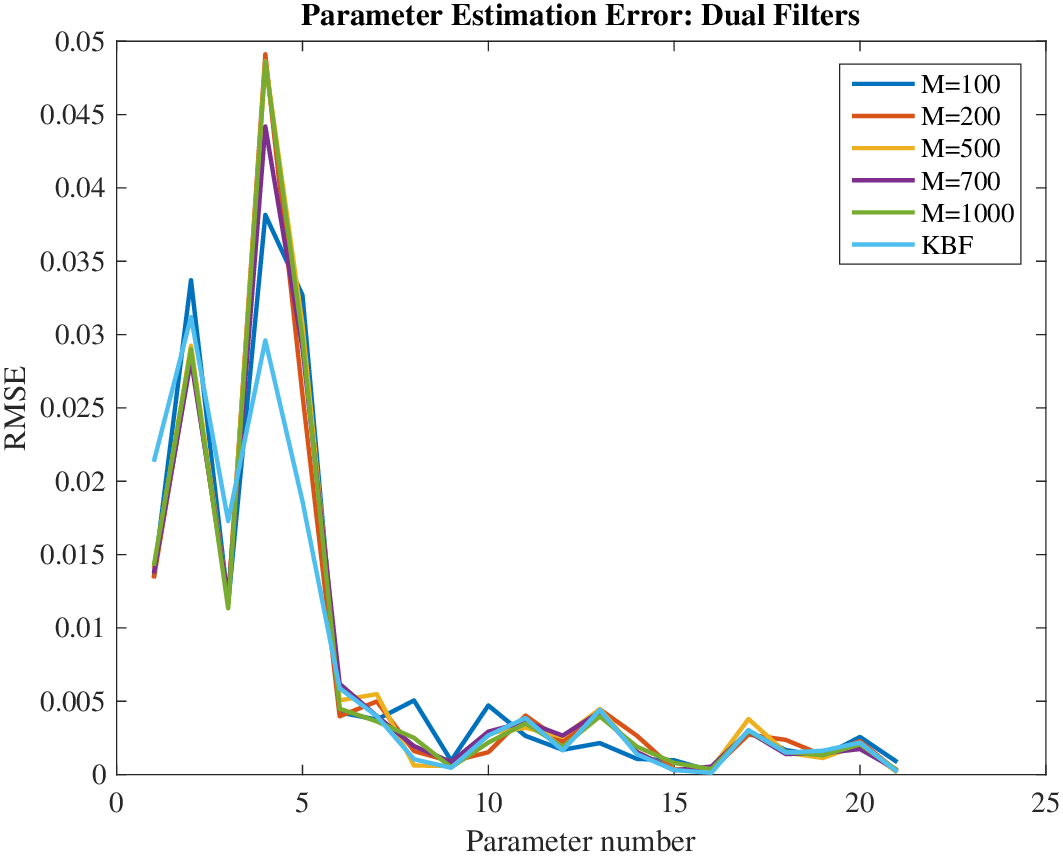}
\label{fig:subfig 7.8a}}
\caption[Plots of the root mean square error and boxplots for the hyper-parameter estimates obtained using KBF-EnKBF and EnKBF: advection equation]{(a) and (b) are, respectively, the box-plots showing the distribution of hyper-parameter estimates of EnKBF dual filter, applied to the advection equation, after a burn-in of $500$ iterations and the plots of the root mean square error for the hyper-parameter estimates obtained using KBF-EnKBF and EnKBF (with different ensemble sizes) dual filters.}
\label{fig:globfig 7.8}
\end{figure}

That the boxplots of EnKBF dual filter parameter estimates have short whiskers (see \Cref{fig:subfig 7.8b}) and few outliers and that the estimates match the true parameter values indicates that EnKBF dual filter is robust in this setting. The EnKBF dual filter registers a slight variation in RMSE from that of the KBF-EnKBF dual filter. This indicates that the EnKBF, in this setting, performs optimally.  

We now repeat the same procedure but with the wave equation described in \Cref{subsec:7.3.2} in the place of the advection equation. The following panels show the results.
\begin{figure}[htp]
\centering
\subfigure[Estimate of parameter $A_{0}$][]{
\includegraphics[width=0.45\textwidth]{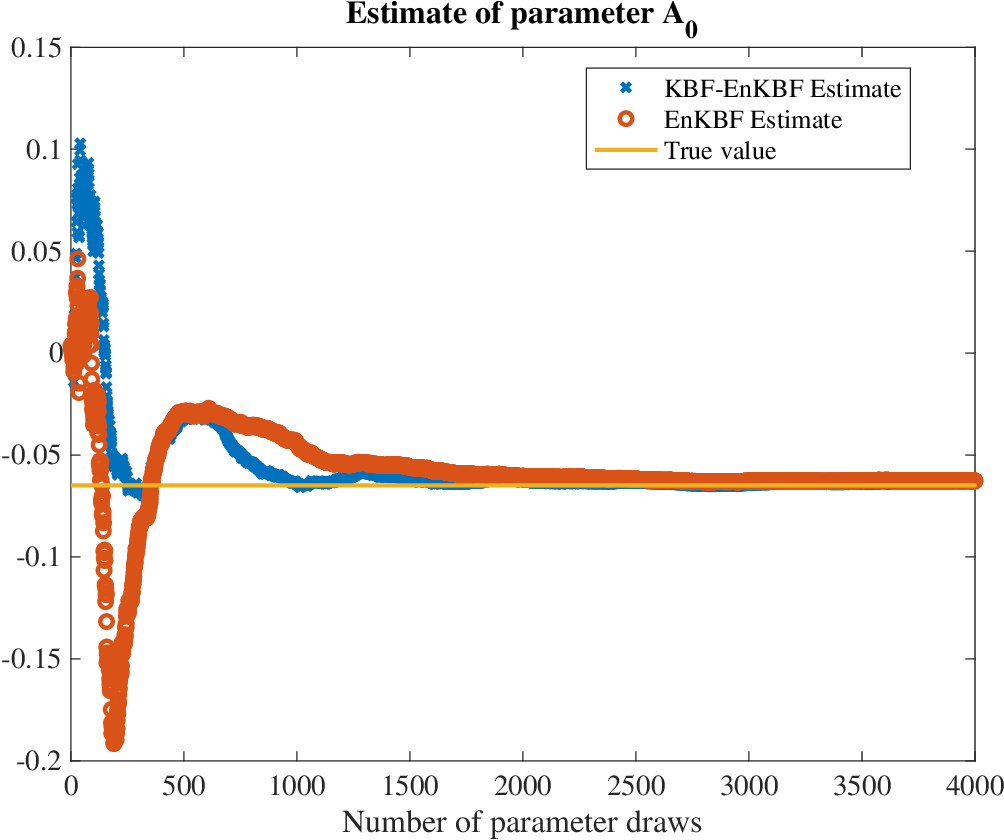}
\label{fig:subfig 5.7a}}
\subfigure[Estimate of parameter $A_{1}$][]{
\includegraphics[width=0.45\textwidth]{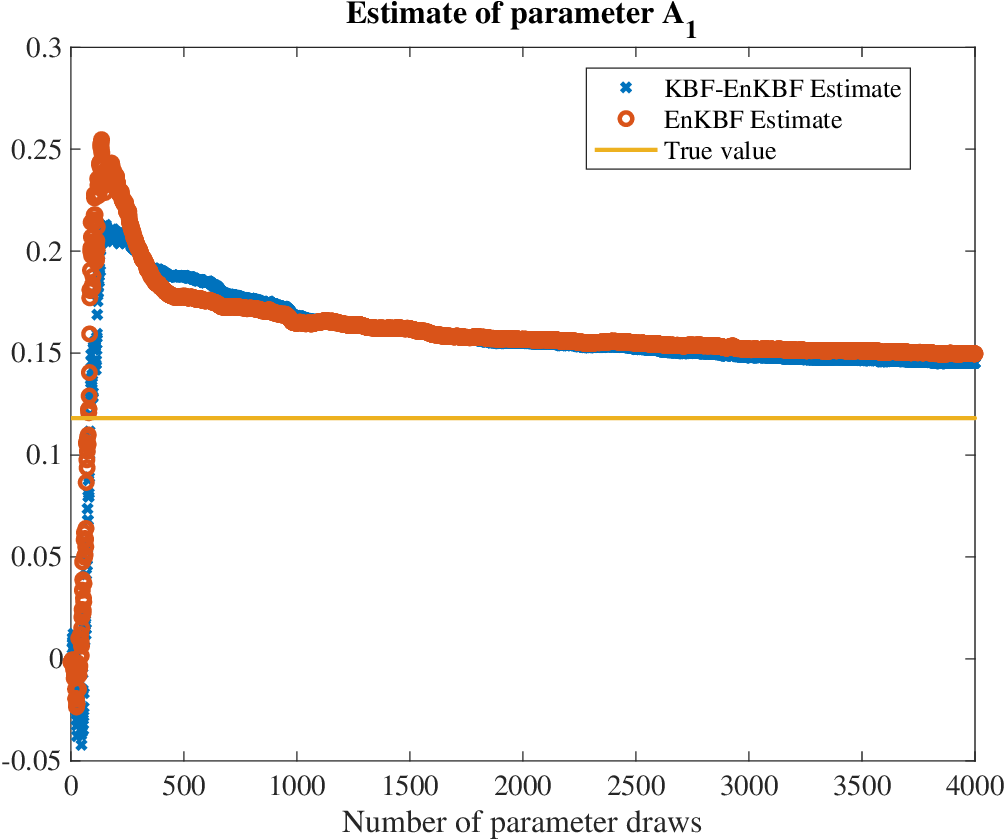}
\label{fig:subfig 5.7b}}
\caption[Plots for velocity parameters in the wave equation for the first two parameters in $\lambda$ obtained using KBF-EnKBF  and EnKBF dual filters, both run for $1000$ time steps]{(a), (b) and (c) are plots for velocity parameters for the first three parameters in $\lambda$ obtained using KBF-EnKBF  and EnKBF dual filters, applied to the wave equation (Example 7.3.1), both filters run for $1000$ time steps. The number of particles, for the state and hyper-parameters, used in EnKBF filter is $M=1000$; the time step size used in both filters is $dt=0.01$. $\mu=0.001$ and $100$ grid points are used. Compared to the results obtained using Metropolis-Hastings algorithm (\Cref{subsec:5.4.1}), the dual filters register a dismal performance, at least in this example.}
\label{fig:globfig 5.7}
\end{figure}
The results shown in \Cref{fig:globfig 5.7} are indicative of a dismal performance of the two dual filters---KBF-EnKBF and EnKBF dual filters---when applied to the wave equation as compared to the results obtained when the dual filters are applied to the advection equation (see \Cref{fig:globfig 7.7}). We note that the wave equation is partially observed; that is, $u$ only is observed in the discretised wave equation, \cref{eq:7.1.13}, whereas the advection equation is fully observed. Furthermore, the number of unknowns in the state-parameter system of the wave equation is $300$ whilst that in the advection equation is $200$. These account for the dismal performance of the KBF-EnKBF and EnKBF dual filters when applied to the wave equation. 

\begin{figure}[H]
\centering
\subfigure[Boxplots of parameter estimates: dual filters][]{
\includegraphics[width=0.45\textwidth]{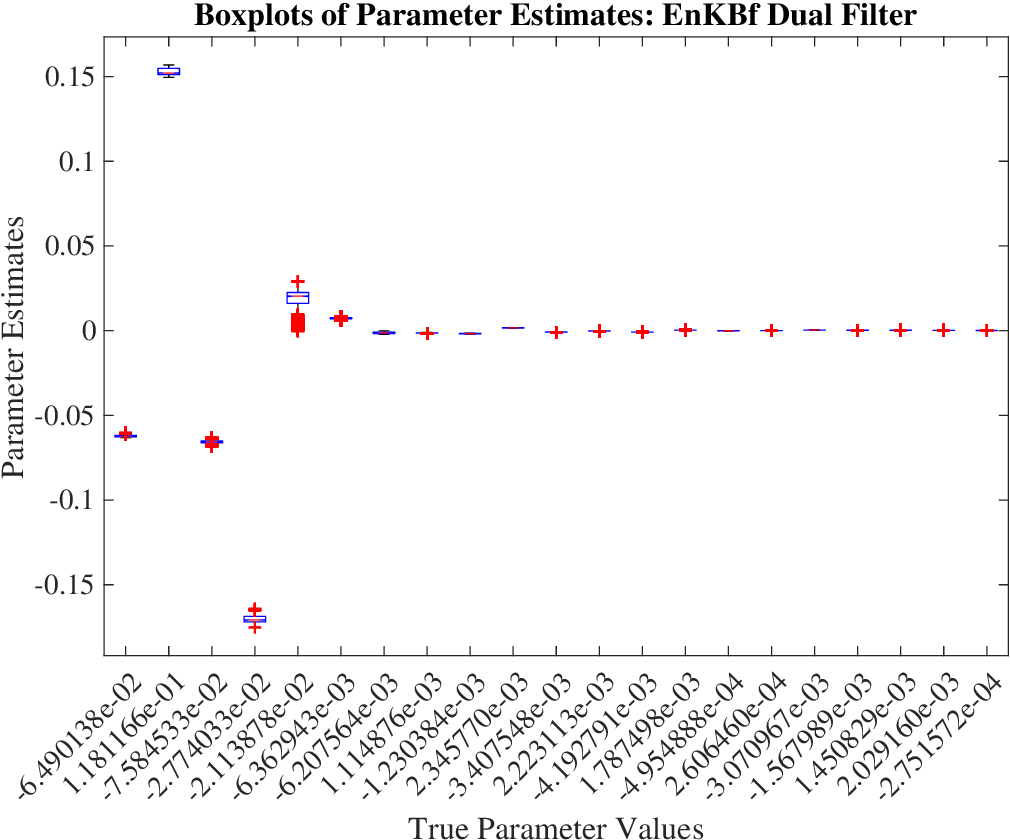}
\label{fig:subfig 5.8b}}
\subfigure[Parameter estimation error: dual filters][]{
\includegraphics[width=0.45\textwidth]{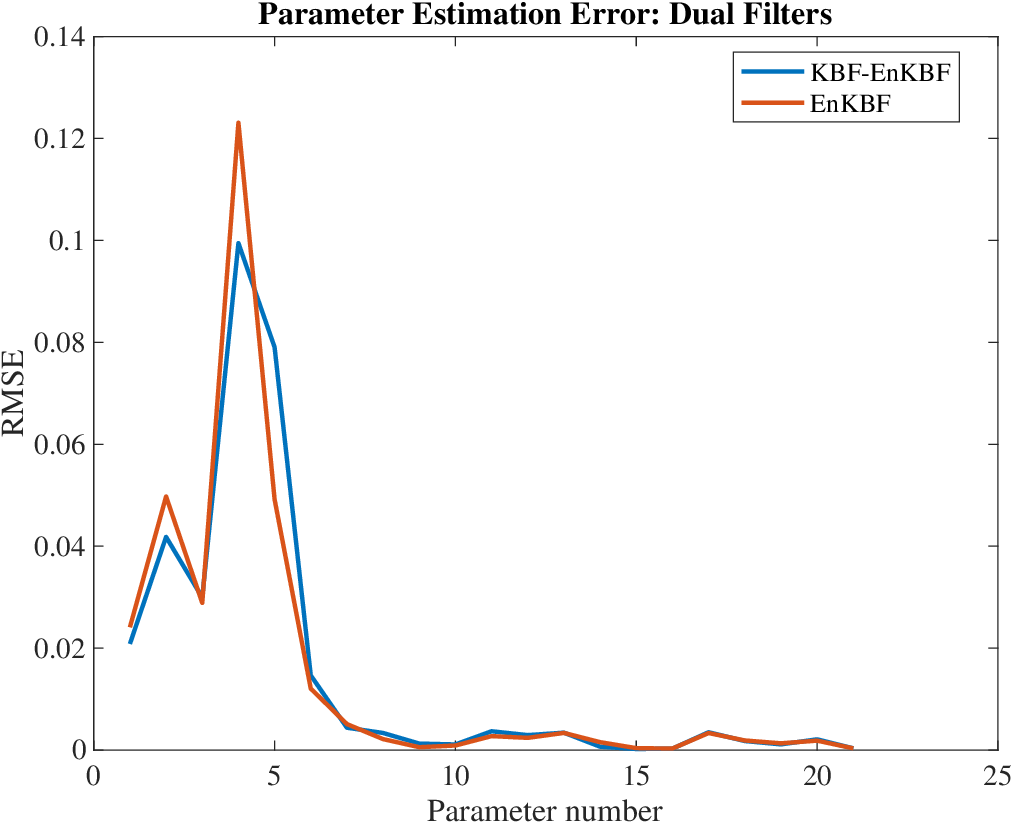}
\label{fig:subfig 5.8a}}
\caption[Plots of the root mean square error and boxplots for the hyper-parameter estimates obtained using KBF-EnKBF and EnKBF: wave equation]{(a) and (b) are, respectively, the box-plots showing the distribution of the $21$ hyper-parameter estimates of EnKBF dual filter after a burn-in of $500$ iterations and plots of the root mean square error for the hyper-parameter estimates obtained using KBF-EnKBF and EnKBF (with different ensemble sizes) dual filters, applied to the wave equation (\Cref{exam:6}).}
\label{fig:globfig 5.8}
\end{figure}

In \Cref{fig:globfig 5.8} are plotted the root mean square errors for parameter estimates for both the EnKBF and KBF and the box-plots showing the dispersion of parameter estimates resulting from the use of EnKBF. 

\section{Conclusions}
\label{sec:conclusions}
We have considered a special case of parameter estimation where the parameter to be estimated is spatially varying. Such a case arises, for example, in the velocity of a wave travelling through an inhomogeneous media. We proposed and studied two approaches: the use of filter likelihood and Metropolis Hastings procedure and joint estimation of state and parameters. The parameter is expressed as a Fourier series with constant coefficients. The coefficients are approximated and then substituted back to the Fourier series to obtain an approximation of the velocity. Kalman-Bucy filter and the ensemble Kalman-Bucy filter are used. The filter likelihood with Metropolis Hastings procedure register a better performance compared to the joint estimation procedure in both advection and wave equations. From the foregoing, Metropolis-Hastings with the filter evidence performs well in estimation of parameters compared to the dual filters---especially when the number of unknowns is large. This is is indicative of the robustness of the Metropolis-Hastings algorithm in searching, and remaining in the, high probability region of the state-space.

\section*{Acknowledgments}
The author would like to acknowledge the assistance of Professor Dr. Sebastian Reich, of Potsdam University, Germany, in supervising my doctoral research \cite{Ang1}, of which this paper is part.

\bibliographystyle{siamplain}
\bibliography{mybib}
\end{document}